\documentclass[twocolumn,times]{aastex631}

\usepackage{multirow}
\usepackage{rotating}
\usepackage{amsmath}
\usepackage{longtable}
\usepackage{booktabs}
\usepackage{ltablex}
\usepackage{epsfig}
\usepackage{threeparttable}
\usepackage{url}
\usepackage{xcolor}
\usepackage{color}
\usepackage{hyperref}
\usepackage{soul}

\def\nustar{\emph{NuSTAR}}

\def\swift{\emph{Swift}}
\def\nicer{\emph{NICER}}
\def\ep{\emph{Einstein Probe}}
\def\hxmt{\emph{Insight}-HXMT}

\def\cm2{\rm \ cm$^{-2}$}

\def\deg{\mbox{$^{\circ}$}}

\newcommand\T{\rule{0pt}{2.6ex}}       
\newcommand\B{\rule[-1.2ex]{0pt}{0pt}}

\shorttitle{Discovery of the black hole X-ray binary candidate EP J182730.0-095633}
\shortauthors{H. Q. Cheng et al.}

\begin{document}

\title{\Large \textit{Einstein Probe} Discovery of EP J182730.0-095633: A New Black Hole X-ray Binary Candidate in Faint Outburst?}

\correspondingauthor{L. Tao}

\author[0000-0003-4200-9954]{H.~Q.~Cheng}
\affiliation{National Astronomical Observatories, Chinese Academy of Sciences, 20A Datun Road, Beijing 100101, China}

\author[0000-0001-9893-8248]{Q.~C.~Zhao}
\affiliation{Key Laboratory of Particle Astrophysics, Institute of High Energy Physics, Chinese Academy of Sciences, Beijing 100049, China}
\affiliation{School of Astronomy and Space Science, University of Chinese Academy of Sciences, 19A Yuquan Road, Beijing 100049, China}

\author[0000-0002-2705-4338]{L.~Tao}
\email{taolian@ihep.ac.cn}
\affiliation{Key Laboratory of Particle Astrophysics, Institute of High Energy Physics, Chinese Academy of Sciences, Beijing 100049, China}

\author[0000-0001-7584-6236]{H.~Feng}
\affiliation{Key Laboratory of Particle Astrophysics, Institute of High Energy Physics, Chinese Academy of Sciences, Beijing 100049, China}

\author[0000-0001-7611-1581]{F.~Coti Zelati}
\affiliation{Institute of Space Sciences (ICE), CSIC, Campus UAB, Barcelona, E-08193, Spain}
\affiliation{ Institut d’Estudis Espacials de Catalunya (IEEC), Barcelona, E-08034, Spain}

\author{H.~W.~Pan}
\affiliation{National Astronomical Observatories, Chinese Academy of Sciences, 20A Datun Road, Beijing 100101, China}

\author{A.~L.~Wang}
\affiliation{Key Laboratory of Particle Astrophysics, Institute of High Energy Physics, Chinese Academy of Sciences, Beijing 100049, China}

\author{Y.~N.~Wang}
\affiliation{National Astronomical Observatories, Chinese Academy of Sciences, 20A Datun Road, Beijing 100101, China}

\author{M.~Y.~Ge}
\affiliation{Key Laboratory of Particle Astrophysics, Institute of High Energy Physics, Chinese Academy of Sciences, Beijing 100049, China}

\author[0000-0001-5990-6243]{A.~Rau}
\affiliation{Max Planck Institute for Extraterrestrial Physics, Garching, 85748, Germany}

\author[0000-0001-5674-4664]{A.~Marino}
\affiliation{Institute of Space Sciences (ICE), CSIC, Campus UAB, Barcelona, E-08193, Spain}
\affiliation{ Institut d’Estudis Espacials de Catalunya (IEEC), Barcelona, E-08034, Spain}

\author{L.~Zhang}
\affiliation{Key Laboratory of Particle Astrophysics, Institute of High Energy Physics, Chinese Academy of Sciences, Beijing 100049, China}

\author[0009-0003-9214-7316]{W.~J.~Zhang}
\affiliation{National Astronomical Observatories, Chinese Academy of Sciences, 20A Datun Road, Beijing 100101, China}

\author[0000-0002-0426-3276]{F. Carotenuto}
\affiliation{INAF-Osservatorio Astronomico di Roma, Via Frascati 33, I-00078, Monte Porzio Catone (RM), Italy}

\author[0000-0001-9599-7285]{L.~Ji}
\affiliation{School of Physics and Astronomy, Sun Yat-Sen University, Zhuhai 519082, China}

\author[0000-0002-2006-1615]{C.~C.~Jin}
\affiliation{National Astronomical Observatories, Chinese Academy of Sciences, 20A Datun Road, Beijing 100101, China}
\affiliation{School of Astronomy and Space Science, University of Chinese Academy of Sciences, 19A Yuquan Road, Beijing 100049, China}
\affiliation{Institute for Frontier in Astronomy and Astrophysics, Beijing Normal University, Beijing 102206, China}

\author{D.~Y.~Li}
\affiliation{National Astronomical Observatories, Chinese Academy of Sciences, 20A Datun Road, Beijing 100101, China}

\author{B.~F.~Liu}
\affiliation{National Astronomical Observatories, Chinese Academy of Sciences, 20A Datun Road, Beijing 100101, China}
\affiliation{School of Astronomy and Space Science, University of Chinese Academy of Sciences, 19A Yuquan Road, Beijing 100049, China}

\author{Y.~Liu}
\affiliation{National Astronomical Observatories, Chinese Academy of Sciences, 20A Datun Road, Beijing 100101, China}

\author{E.~L.~Qiao}
\affiliation{National Astronomical Observatories, Chinese Academy of Sciences, 20A Datun Road, Beijing 100101, China}
\affiliation{School of Astronomy and Space Science, University of Chinese Academy of Sciences, 19A Yuquan Road, Beijing 100049, China}

\author[0000-0003-2177-6388]{N.~Rea}
\affiliation{Institute of Space Sciences (ICE), CSIC, Campus UAB, Barcelona, E-08193, Spain}
\affiliation{ Institut d’Estudis Espacials de Catalunya (IEEC), Barcelona, E-08034, Spain}

\author[0000-0002-4622-796X]{R.~Soria}
\affiliation{INAF – Osservatorio Astrofisico di Torino, Pino Torinese, I-10025, Italy}
\affiliation{Sydney Institute for Astronomy, School of Physics A28, The University of Sydney, NSW 2006, Australia}

\author{S.~Wang}
\affiliation{National Astronomical Observatories, Chinese Academy of Sciences, 20A Datun Road, Beijing 100101, China}

\author[0000-0002-5385-9586]{Z.~Yan}
\affiliation{Shanghai Astronomical Observatory, Chinese Academy of Sciences, 80 Nandan Road, Shanghai, 200030, China}

\author{W.~Yuan}
\affiliation{National Astronomical Observatories, Chinese Academy of Sciences, 20A Datun Road, Beijing 100101, China}
\affiliation{School of Astronomy and Space Science, University of Chinese Academy of Sciences, 19A Yuquan Road, Beijing 100049, China}

\author[0000-0002-9725-2524]{B.~Zhang}
\affiliation{Nevada Center for Astrophysics, University of Nevada Las Vegas, NV 89154, USA}
\affiliation{Department of Physics and Astronomy, University of Nevada Las Vegas, NV 89154, USA}

\author[0000-0001-8630-5435]{G.~B.~Zhang}
\affiliation{Yunnan Observatories, Chinese Academy of Sciences, Kunming 650216, China}
\affiliation{Key Laboratory for the Structure and Evolution of Celestial Objects, Chinese Academy of Sciences, Kunming 650216, China}

\author{S.~N.~Zhang}
\affiliation{Key Laboratory of Particle Astrophysics, Institute of High Energy Physics, Chinese Academy of Sciences, Beijing 100049, China}
\affiliation{School of Astronomy and Space Science, University of Chinese Academy of Sciences, 19A Yuquan Road, Beijing 100049, China}

\author{W.~D.~Zhang}
\affiliation{National Astronomical Observatories, Chinese Academy of Sciences, 20A Datun Road, Beijing 100101, China}


\author{A.~Beardmore}
\affiliation{School of Physics and Astronomy, University of Leicester, University Road, Leicester, LE1 7RH, UK}

\author[0000-0002-7735-5796]{J. S. Bright}
\affiliation{Astrophysics, Department of Physics, University of Oxford, Keble Road, Oxford OX1 3RH, UK}

\author[0009-0000-4068-1320]{X.~L.~Chen}
\affiliation{South-Western Institute for Astronomy Research, Yunnan University, Kunming, Yunnan 650504, China}

\author{Z.~Fan}
\affiliation{National Astronomical Observatories, Chinese Academy of Sciences, 20A Datun Road, Beijing 100101, China}
\affiliation{School of Astronomy and Space Science, University of Chinese Academy of Sciences, 19A Yuquan Road, Beijing 100049, China}

\author{S.~Y.~Fu}
\affiliation{National Astronomical Observatories, Chinese Academy of Sciences, 20A Datun Road, Beijing 100101, China}

\author[0000-0002-8149-8298]{J.~P.~U.~Fynbo}
\affiliation{Cosmic Dawn Center (DAWN)}
\affiliation{Niels Bohr Institute, University of Copenhagen, Jagtvej 128, 2200, Copenhagen N, Denmark}

\author[0000-0002-0779-1947]{J.~W.~Hu}
\affiliation{National Astronomical Observatories, Chinese Academy of Sciences, 20A Datun Road, Beijing 100101, China}

\author{J.~J.~Jin}
\affiliation{National Astronomical Observatories, Chinese Academy of Sciences, 20A Datun Road, Beijing 100101, China}

\author[0000-0001-5679-0695]{P. G. Jonker}
\affiliation{Department of Astrophysics/IMAPP, Radboud University, P.O. Box 9010, 6500 GL, Nijmegen, The Netherlands}

\author{A.~K.~H.~Kong}
\affiliation{Institute for Cosmic Ray Research, The University of Tokyo, Kashiwa City, Chiba 277-8582, Japan}

\author{E. Kuulkers}
\affiliation{ESA/ESTEC, Noordwijk, 2201 AZ, the Netherlands}

\author{C.~K.~Li}
\affiliation{Key Laboratory of Particle Astrophysics, Institute of High Energy Physics, Chinese Academy of Sciences, Beijing 100049, China}

\author{H.~L.~Li}
\affiliation{National Astronomical Observatories, Chinese Academy of Sciences, 20A Datun Road, Beijing 100101, China}

\author{Z.~K.~Lin}
\affiliation{National Astronomical Observatories, Chinese Academy of Sciences, 20A Datun Road, Beijing 100101, China}

\author[0000-0001-5561-2010]{C.~X.~Liu}
\affiliation{South-Western Institute for Astronomy Research, Yunnan University, Kunming, Yunnan 650504, China}

\author{H.-Y.~Liu}
\affiliation{National Astronomical Observatories, Chinese Academy of Sciences, 20A Datun Road, Beijing 100101, China}

\author[0000-0002-7420-6744]{J.~Z.~Liu}
\affiliation{Xinjiang Astronomical Observatory, Chinese Academy of Sciences, Urumqi 830011, China}
\affiliation{School of Astronomy and Space Science, University of Chinese Academy of Sciences, 19A Yuquan Road, Beijing 100049, China}

\author[0000-0003-1295-2909]{X.~W.~Liu}
\affiliation{South-Western Institute for Astronomy Research, Yunnan University, Kunming, Yunnan 650504, China}

\author{Z.~Lu}
\affiliation{College of Mathematics and Physics, China Three Gorges University, Yichang 443002, China}
\affiliation{Center for Astronomy and Space Sciences, China Three Gorges University, Yichang 443002, China}

\author{C.~Maitra}
\affiliation{Max Planck Institute for Extraterrestrial Physics, Garching, 85748, Germany}

\author{H.~Y.~Mu}
\affiliation{National Astronomical Observatories, Chinese Academy of Sciences, 20A Datun Road, Beijing 100101, China}

\author[0000-0002-5847-2612]{C.-Y.~Ng}
\affiliation{Department of Physics, University of Hong Kong, Pokfulam Road, Hong Kong, China}

\author{Y.~L.~Qiu}
\affiliation{National Astronomical Observatories, Chinese Academy of Sciences, 20A Datun Road, Beijing 100101, China}

\author[0000-0002-1481-4676]{S.~Tinyanont}
\affiliation{National Astronomical Research Institute of Thailand, 260 Moo 4, Donkaew, Maerim, Chiang Mai, 50180, Thailand}

\author[0000-0002-8385-7848]{Y.~Wang}
\affiliation{Key Laboratory of Dark Matter and Space Astronomy, Purple Mountain Observatory, Chinese Academy of Sciences, Nanjing 210023, China}

\author{S.~X.~Wen}
\affiliation{National Astronomical Observatories, Chinese Academy of Sciences, 20A Datun Road, Beijing 100101, China}

\author{S.~S.~Weng}
\affiliation{Department of Physics and Institute of Theoretical Physics, Nanjing Normal University, Nanjing, China}

\author{Jianfeng Wu}
\affiliation{ Department of Astronomy, Xiamen University, Xiamen, Fujian 361005, China}

\author{D.~Xu}
\affiliation{National Astronomical Observatories, Chinese Academy of Sciences, 20A Datun Road, Beijing 100101, China}

\author{Y.~K.~Yan}
\affiliation{College of Mathematics and Physics, China Three Gorges University, Yichang 443002, China}

\author{Z.~Yan}
\affiliation{Yunnan Observatories, Chinese Academy of Sciences, Kunming 650216, China}

\author[0000-0001-6374-8313]{Y.-P.~Yang}
\affiliation{South-Western Institute for Astronomy Research, Yunnan University, Kunming, Yunnan 650504, China}

\author{P.~Zhang}
\affiliation{College of Mathematics and Physics, China Three Gorges University, Yichang 443002, China}
\affiliation{Center for Astronomy and Space Sciences, China Three Gorges University, Yichang 443002, China}

\author{S.~Zhang}
\affiliation{Key Laboratory of Particle Astrophysics, Institute of High Energy Physics, Chinese Academy of Sciences, Beijing 100049, China}

\author{Q.~Zhao}
\affiliation{National Astronomical Observatories, Chinese Academy of Sciences, 20A Datun Road, Beijing 100101, China}


\author{Z.~M.~Cai}
\affiliation{Innovation Academy for Microsatellites, Chinese Academy of Sciences, Shanghai 201210, China}

\author{Y.~Chen}
\affiliation{Key Laboratory of Particle Astrophysics, Institute of High Energy Physics, Chinese Academy of Sciences, Beijing 100049, China}

\author{Y.~F.~Chen}
\affiliation{Shanghai Institute of Technical Physics, Chinese Academy of Sciences, Shanghai 200083, China}

\author{C.~Z.~Cui}
\affiliation{National Astronomical Observatories, Chinese Academy of Sciences, 20A Datun Road, Beijing 100101, China}
\affiliation{School of Astronomy and Space Science, University of Chinese Academy of Sciences, 19A Yuquan Road, Beijing 100049, China}

\author{W.~W.~Cui}
\affiliation{Key Laboratory of Particle Astrophysics, Institute of High Energy Physics, Chinese Academy of Sciences, Beijing 100049, China}

\author{H.~B.~Hu}
\affiliation{National Astronomical Observatories, Chinese Academy of Sciences, 20A Datun Road, Beijing 100101, China}

\author{M.~H.~Huang}
\affiliation{National Astronomical Observatories, Chinese Academy of Sciences, 20A Datun Road, Beijing 100101, China}
\affiliation{School of Astronomy and Space Science, University of Chinese Academy of Sciences, 19A Yuquan Road, Beijing 100049, China}

\author{S.~M.~Jia}
\affiliation{Key Laboratory of Particle Astrophysics, Institute of High Energy Physics, Chinese Academy of Sciences, Beijing 100049, China}

\author{G.~Jin}
\affiliation{North Night Vision Technology Co., LTD, Nanjing, China}

\author{Z.~X.~Ling}
\affiliation{National Astronomical Observatories, Chinese Academy of Sciences, 20A Datun Road, Beijing 100101, China}
\affiliation{School of Astronomy and Space Science, University of Chinese Academy of Sciences, 19A Yuquan Road, Beijing 100049, China}
\affiliation{Institute for Frontier in Astronomy and Astrophysics, Beijing Normal University, Beijing 102206, China}

\author{H.~Q.~Liu}
\affiliation{Innovation Academy for Microsatellites, Chinese Academy of Sciences, Shanghai 201210, China}

\author{S.~L.~Sun}
\affiliation{Shanghai Institute of Technical Physics, Chinese Academy of Sciences, Shanghai 200083, China}

\author{X.~J.~Sun}
\affiliation{Shanghai Institute of Technical Physics, Chinese Academy of Sciences, Shanghai 200083, China}

\author{Y.~F.~Xu}
\affiliation{National Astronomical Observatories, Chinese Academy of Sciences, 20A Datun Road, Beijing 100101, China}
\affiliation{School of Astronomy and Space Science, University of Chinese Academy of Sciences, 19A Yuquan Road, Beijing 100049, China}

\author{C.~Zhang}
\affiliation{National Astronomical Observatories, Chinese Academy of Sciences, 20A Datun Road, Beijing 100101, China}
\affiliation{School of Astronomy and Space Science, University of Chinese Academy of Sciences, 19A Yuquan Road, Beijing 100049, China}

\author{M.~Zhang}
\affiliation{National Astronomical Observatories, Chinese Academy of Sciences, 20A Datun Road, Beijing 100101, China}

\author{Y.~H.~Zhang}
\affiliation{Innovation Academy for Microsatellites, Chinese Academy of Sciences, Shanghai 201210, China}


\received{}
\revised{}
\accepted{}
\submitjournal{ApJL}

\begin{abstract}

Black hole X-ray binaries (candidates) currently identified in our galaxy are mainly transient sources, with the majority discovered through the detection of their X-ray outbursts. Among these, only four were found during faint outbursts exhibiting peak X-ray luminosities $L_{\rm X}\lesssim10^{36}~{\rm erg~s^{-1}}$, likely due to the previous lack of sensitive, wide-field monitoring instruments in the X-ray band.
In this \textit{Letter}, we present the discovery of an intriguing X-ray transient, EP J182730.0-095633, via the \textit{Einstein Probe} (\textit{EP}) and subsequent multi-wavelength follow-up studies.
This transient, located on the Galactic plane, experienced a faint and brief X-ray outburst lasting about 20 days.
Its X-ray spectrum is non-thermal and consistent with a power-law model with a nearly constant photon index of $\Gamma \sim2$ throughout the outburst. 
A long-lasting millihertz quasi-periodic oscillation (QPO) signal was detected in its X-ray light curve, centered around a frequency of $\sim0.04$ Hz. 
A transient near-infrared source was identified as its counterpart, although no optical emission was detectable, likely due to significant extinction.
A radio counterpart was also observed, displaying an inverted radio spectrum with $\alpha\sim0.45$.
The X-ray spectral and temporal characteristics, along with the multi-wavelength properties, indicate that the source is a faint low-mass X-ray binary, with the compact object likely being a black hole. 
This work demonstrates the potential of the \textit{EP} in discovering new X-ray binaries by capturing faint-level X-ray outbursts.  

\end{abstract}



\section{Introduction} \label{sec:intro}

Black holes (BHs) are among the most mysterious celestial objects in the universe. 
They provide a unique laboratory for studying the General Relativity effects and other physical processes in strong gravitational fields. 
As the inevitable evolutionary destiny and remnant of massive stars, 
stellar-mass BHs, with masses in the typical range of several to several tens of solar masses, 
are of great importance in understanding the formation and growth of BHs, as well as the evolution of massive stars. 
So far, stellar-mass BHs have been found to exist
in binary systems termed X-ray binaries \citep[XRBs, e.g.][]{vanParadijs1995,Lewin1995,Remillard_review}. 
In our galaxy, they have preferably been found in binary systems with a low-mass companion star, termed Low-Mass X-ray Binaries \citep[LMXBs, see][for detailed reviews]{Kalemci_review,Bahramian_LMXB}. 
LMXBs are mostly X-ray transient sources, exhibiting bright, sporadic outbursts that typically last from months to years \citep[e.g.][]{Remillard_review, Belloni_REVIEW}.

Currently, over 70 transient BH (candidate) X-ray binaries have been identified\footnote{\url{https://www.astro.puc.cl/BlackCAT/transients.php}}. Among these, about $20$ contain a compact object confirmed to be in the BH mass range using dynamical methods, i.e., by measuring the orbital motion of their companion stars \citep{blackcat, XRBcats}. 
Historically, these systems were predominantly discovered via capturing their transient outbursts 
by a series of wide-field X-ray monitoring instruments that have been flown since the 1970's, such as the \textit{Ariel-V} \citep{Villa1976}, the all-sky monitor WATCH on board Granat satellite \citep{Brandt1990},
the All-Sky Monitor (ASM) on board \textit{Rossi X-ray Timing Explorer} \citep[\textit{RXTE};][]{LevineRXTE1996}, the INTErnational Gamma-Ray Astrophysics Laboratory \citep[\textit{INTEGRAL};][]{Winkler2003, Kuulkers2021}, the Burst Alert Telescope \citep[BAT;][]{BAT2005} on board the \textit{Neil Gehrels Swift Observatory} \citep[\textit{Swift};][]{Gehrels_Swift}, and the Monitor of All-sky X-ray Image \citep[\textit{MAXI;}][]{MAXI2009}.
For the vast majority of these systems, the detected X-ray outbursts reached a peak luminosity typically over $10^{37}~{\rm erg~s^{-1}}$ \citep[e.g.][]{Yan2015, watchdog},  making them bright enough to be detected by most of the X-ray monitors mentioned above.

Only four BHs (candidates) were discovered by the detection of their X-ray emission during faint outbursts, with a peak luminosity $L_{\rm X}\lesssim10^{36}~{\rm erg~s^{-1}}$ \citep[][]{Wijnands_2006}, namely, Swift J1357.2–0933 \citep{CorralSantana2013Science}, XTE J1118+480 \citep{Wagner2001}, XTE J1728-295 \citep{Swank2001, Walter2004}, and CXOGC J174540.0-290031 \citep{Muno2005a, Porquet2005}. 
The first two systems were confirmed to harbor a black hole through dynamical mass measurements, while the latter two were well consistent with BH transients in their observed properties.
The small number of BHs found during faint outbursts can be explained in the sense that the peak X-ray brightness of the majority of such systems fall below the sensitivity of the past and most current wide-field monitors, while the sensitive X-ray telescopes (such as \textit{Chandra} and \textit{XMM-Newton}) have too small field of view (FoV) to catch the outbursts in real time (the case of CXOGC J174540.0-290031 is an exception and the chance of such detections is expected to be low).

It has long been recognized that the number of black hole X-ray binaries (BHXRBs) known so far is much smaller than what was hypothesized to exist in our galaxy \citep{vandenHeuvel1992}.
Based on the distances and spatial distributions of the known BHXRBs, it was estimated that there are more than $10^{3}$ BHXRBs in our galaxy \citep{blackcat}.
In fact, stellar evolution models predicted that there could be as many as $10^{8}-10^{9}$ BHs lurking in our galaxy \citep[e.g.][]{Brown1994BH, Timmes1996BH, Meier2012BHbook}.
This `missing' population of stellar-mass BHs has been a topic of extensive debate for decades \citep{Hailey2018Nature}.
The apparently very low detectability of BHXRBs is likely attributable to their presumably transient nature and the fact that most of their outbursts are likely too faint to be detectable by the past and most of the current X-ray wide-field monitors. 
It is thus expected that, with the advent of X-ray wide-field monitors having significantly improved sensitivity, more of such systems could be revealed.
Nonetheless, further challenges lie with the identification of the nature of the compact object (be it a BH or a neutron star) for LMXBs, which requires dense monitoring and prompt multi-wavelength follow-up observations \citep{Wijnands_2006, Bahramian_LMXB}.

The \textit{Einstein Probe} \citep[\textit{EP};][]{Yuan22}, launched on January 9th 2024, carries the Wide-field X-ray Telescope (WXT)---a wide-field monitor with a FoV of $\sim3800$ square degrees built from lobster-eye micro-pore optics. 
Operating in the 0.5--4 keV band, WXT can reach a sensitivity of $\sim 1$ mCrab ($\sim (2-3) \times 10^{-11}~{\rm erg~s^{-1}~cm^{-2}}$ in 0.5--4 keV) at an exposure of $\sim1000$ s, which is several-tens times more sensitive than the other currently operating wide-field monitors. 
For BHXRBs within our galaxy, an X-ray outburst brighter than $\sim10^{35}~{\rm erg~s^{-1}}$ will likely be detected by WXT within $\sim10^3~{\rm s}$,
enabling the detection of outbursts that are $\sim 100$ times fainter than those previously detectable. 
The other instrument on board \textit{EP} is the Follow-up X-ray Telescope \citep[FXT,][]{Chen2020FXT} operating in 0.3--10 keV band, which provides source localization of $5-10''$ and sensitivity of $\sim10^{-14}~{\rm erg~s^{-1}~cm^{-2}}$ at an exposure of $10^4~{\rm s}$.
It is expected that \textit{EP} will detect new BHXRB systems, particularly those undergoing faint, short-lived outbursts \citep{2025EP}.

In this \textit{Letter}, we report on the discovery and multi-wavelength follow-up observations of an X-ray transient EP J182730.0-095633 \citep[EP240904a,][]{2024ATel16805....1C} located on the Galactic plane. The spectral, temporal, and multi-wavelength properties observed favor this source being a new BHXRB candidate experiencing a faint outburst lasting for about 20 days.
This paper is organized as follows: in Section~\ref{sec:data_reduction}, we describe the observations and data reduction procedures. The results are presented in Section~\ref{sec:results} and discussed in Section~\ref{sec:discussion}. The conclusion is given in Section~\ref{sec:conclusions}.



\section{Observations and data reduction}
\label{sec:data_reduction}

\subsection{X-ray}\label{sec:xray_datareduction}
\begin{figure*}
    \centering
    \includegraphics[width=1.0\textwidth]{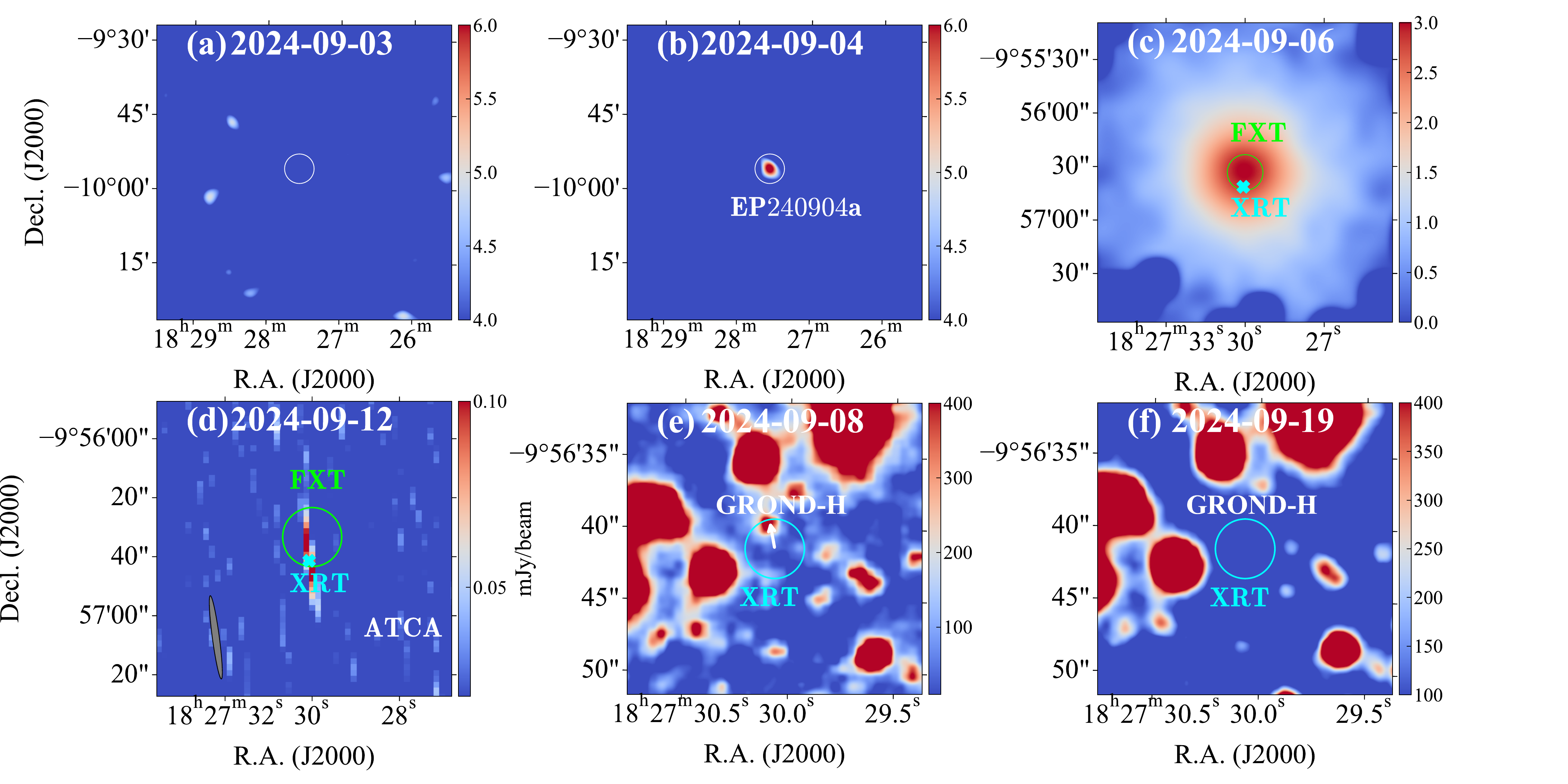}
    \caption{Panels (a) and (b) show \textit{EP}/WXT images of EP J182730.0-095633 from September 3rd to 4th, 2024 (ObsIDs are 06800000068 and 06800000067, respectively). Panel (c) presents \textit{EP}/FXT module A (FXTA) image obtained with Full Frame (FF) mode on September 6th, 2024. Panel (d) displays \textit{ATCA} (5.5\,GHz) images on September 12th, 2024. The restoring beam ($19.5''\times1.2''$, $5.5^{\circ}$) is indicated as a gray elliptical shape in the bottom left corner. Panels (e) and (f) show near-infrared ($H$ bands) images obtained from GROND on September 8th and 19th, 2024, respectively. The potential NIR counterpart is denoted by the white arrow in Panel (e). The white, green, and cyan circles represent the positional errors of WXT, FXT, and XRT, respectively.}
    \label{fig:ep_detection_image}
\end{figure*}

EP J182730.0-095633 was first detected as an X-ray transient (designated EP240904a) in an observation starting on September 4th, 2024, at 10:12:56 UT by \textit{EP}/WXT \citep{2024ATel16805....1C}. 
The source was not detected during a WXT observation taken one day before (Figure~\ref{fig:ep_detection_image}). 
Following the detection of the source, 
a series of follow-up observations were carried out with \textit{EP}/FXT starting on September 6th, 2024 and lasting until September 24th. 
These observations provided a more precise source position at R.A. (J2000.0) = $18^{\rm h}27^{\rm m}30.0^{\rm s}$ and Decl. (J2000.0) = $-9^{\circ}56'33''$, with an uncertainty of $10''$ at the 90\% confidence level \citep{2024ATel16805....1C}.
Meanwhile, several other X-ray telescopes were triggered and joined the follow-up campaign, including the \swift, the \textit{Neutron Star Interior Composition Explorer} \citep[\nicer;][]{Gendreau_2016}, the \textit{Nuclear Spectroscopic Telescope Array} \citep[\nustar;][]{Harrison_nustar}, and the \textit{Hard X-ray Modulation Telescope} \citep[\textit{Insight}-HXMT;][]{Insight_HXMT_Zhang}\footnote{Due to contamination from nearby sources, data from \textit{Insight}-HXMT were excluded from further analysis.}.
The X-ray Telescope (XRT) aboard \swift ~provided an enhanced positional accuracy \citep{Evans2009, Evans2020}, by applying a correction based on the photometry provided by the Ultra-Violet/Optical Telescope (UVOT): R.A. (J2000.0) = $18^{\rm h}27^{\rm m}30.1^{\rm s}$ and Decl. (J2000.0) = $-9^{\circ}56'41.4''$, with an uncertainty of $2.1''$ at the 90\% confidence level. For this work, we adopt the \swift/XRT coordinates as the X-ray position of EP J182730.0-095633. 
It is important to point out that EP J182730.0-095633 resides on the Galactic plane ($l=21.3634$\deg, $b=0.7244$\deg), suggesting a likely Galactic origin for this transient.
The observation log is summarized in Table~\ref{tab:xray_spectralfit}, and the data reduction process is detailed in Section~\ref{sec:detailed_xraydatareduction}.

\subsection{Optical and infrared}
\label{sec:optical_infrared}
Follow-up observations of EP J182730.0-095633 were carried out in the optical and near-infrared bands with various ground- and space-based telescopes. 
These include the Thai Robotic Telescope (TRT), the Nordic Optical Telescope (NOT), the Altay 1-Meter Telescope (ALT100C), the Visible Telescope on board the \textit{Space-based multi-band astronomical Variable Objects Monitor} (\textit{SVOM}/VT), the Tsinghua-NAOC Telescope \citep[TNT,][]{Huang2012TNT} in Xinglong Observatory, the Multi-channel Photometric Survey Telescope \citep[Mephisto,][]{2020Mephisto} and the Gamma-Ray Burst Optical/Near-Infrared Detector \citep[GROND,][]{GROND2008} at the MPG 2.2\,m telescope at ESO's La Silla Observatory. The observational log is listed in Table \ref{tab:EP J182730.0-095633_opt_infrared_followups} and the data reduction of these observations is detailed in Section \ref{sec:optical_datareduction}.

\subsection{Radio}\label{sec:radio}

Follow-up observations of EP J182730.0-095633 were conducted in the radio band with two ground telescopes, \textit{ATCA} and \textit{MeerKAT} \citep{Camilo2018, Jonas2018}. Two \textit{ATCA} observations were carried out on September 12th \citep{2024ATCA_ATEL} and September 21st, 2024. \textit{MeerKAT} observed the source for three times, on September 21st, October 21st and October 28th, 2024. 
The observational log is listed in Table \ref{tab:EP J182730.0-095633_radio_followups} and the data reduction of these observations is detailed in Section \ref{sec:radio_followup}.

\section{Data analysis and results}
\label{sec:results}
\subsection{X-ray spectral analysis}
\label{sec:spectral}

\begin{figure}
    \centering
    \includegraphics[width=0.47\textwidth]{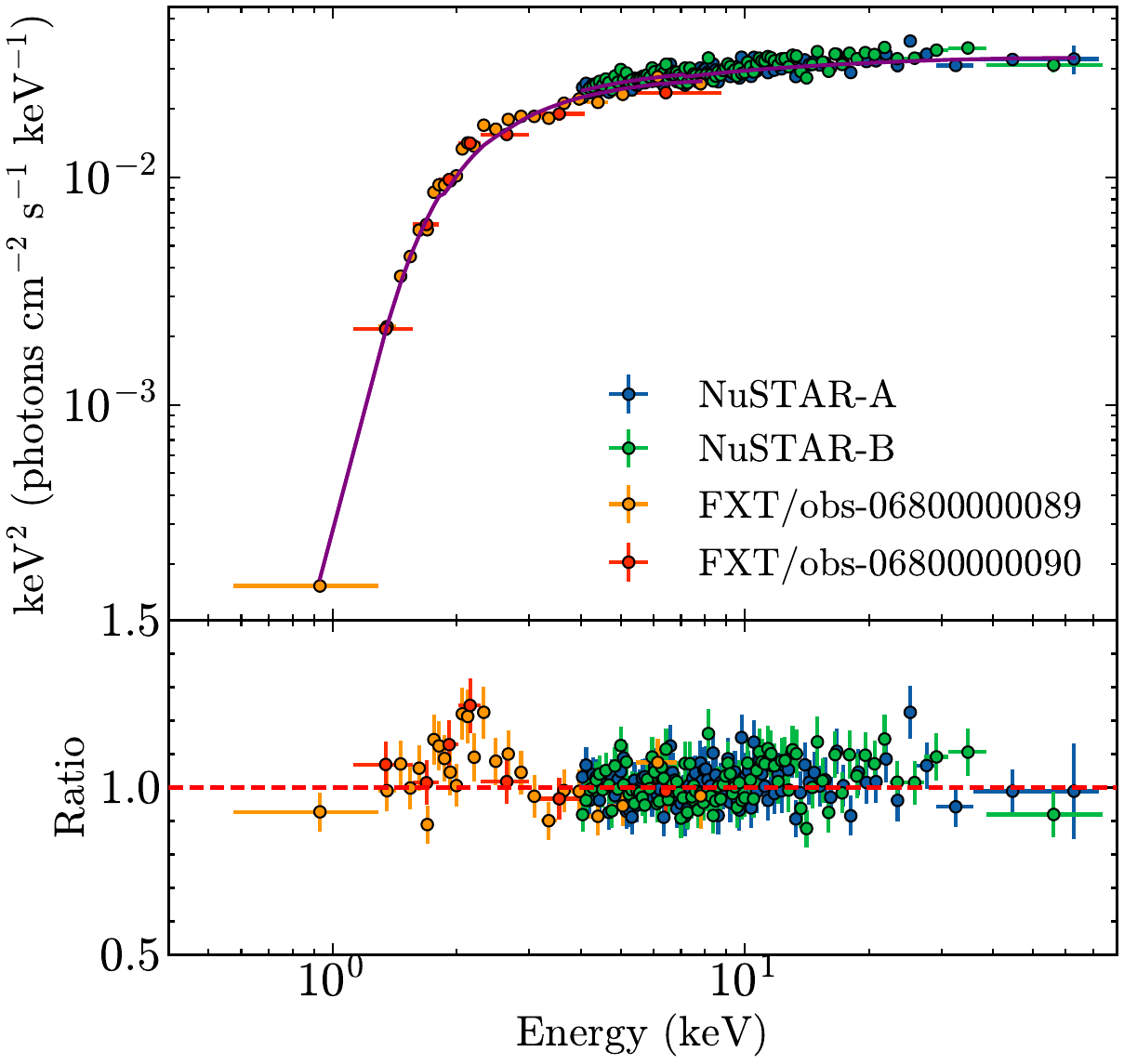}
    \caption{Joint spectral fitting of \textit{EP}/FXT (module B, 0.5--10 keV) and \nustar~(4--79 keV) with an absorbed cutoff power-law model. Note that the residual feature at $\sim2~{\rm keV}$ is due to a calibration issue.}
    \label{fig:spectral_fit}
\end{figure}

\begin{figure*}
    \centering
    \includegraphics[width=0.85\textwidth]{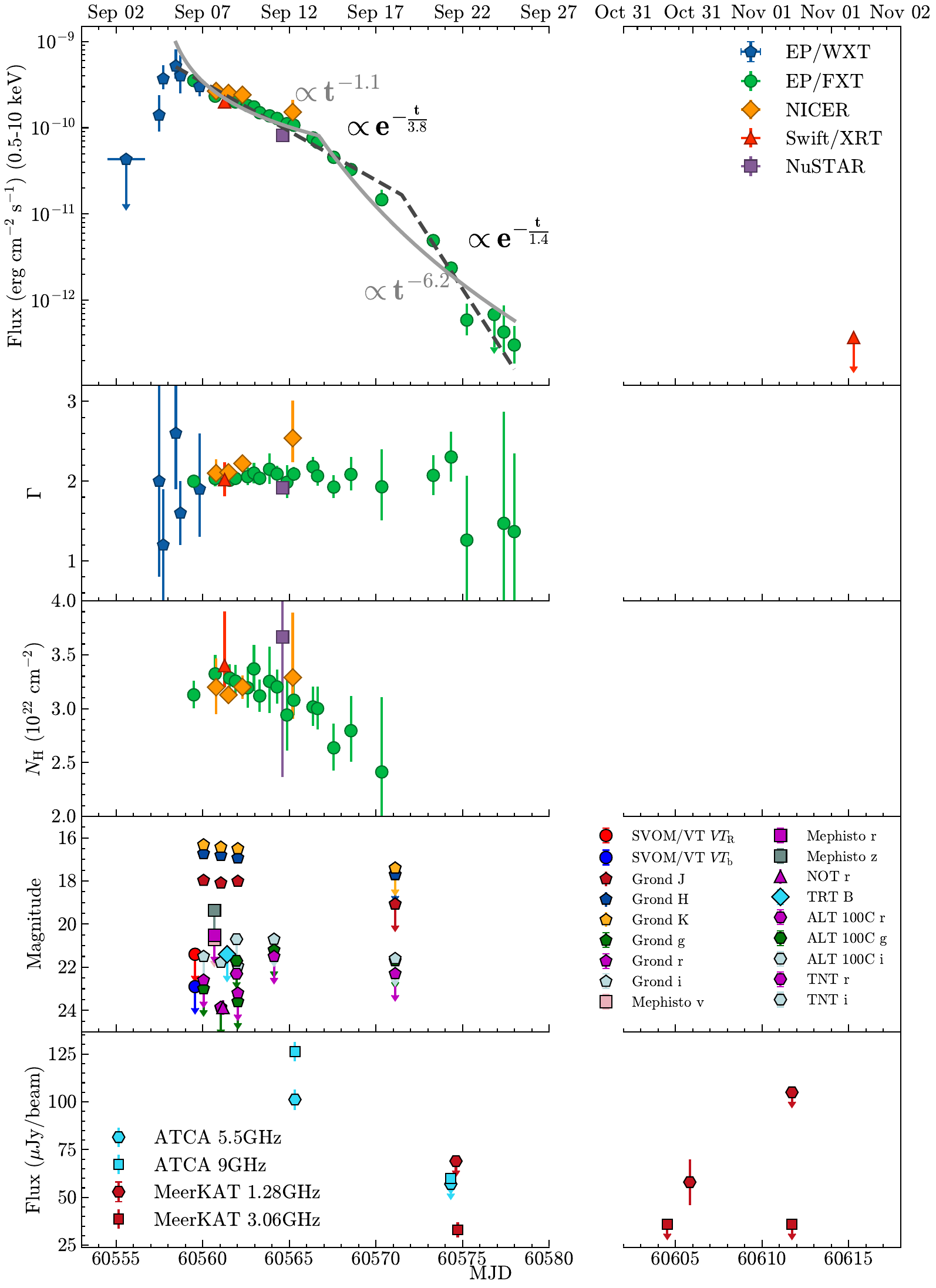}
    \caption{Long-term evolution of the multi-wavelength properties of EP J182730.0-095633. Top to bottom panels: the evolutions of the unabsorbed flux in the 0.5--10 keV band, the photon index ($\Gamma$), the hydrogen column density ($N_{\rm H}$), the observed magnitude in the optical/infrared bands, and the radio peak flux density of \textit{ATCA} and \textit{MeerKAT}, respectively. The X-ray flux during the decaying phase is fitted either by a broken power-law model (the gray solid curve), or a broken exponential model (the black dashed curve). See Section \ref{sec:xray_curve} for more details. The results of a few optical/infrared observations after the X-ray outburst, where only upper limits have been obtained, are not shown for clearer data visualization. Also, primarily due to the low S/N of the data, the X-ray absorbing column density $N_{\rm H}$ in some observations cannot be well constrained within physically reasonable ranges and are fixed at $3.1 \times 10^{22}~{\rm cm}^{-2}$, hence they are not shown in this figure (see text and Table \ref{tab:xray_spectralfit} for more details).}
    \label{fig:xray_evolution}
\end{figure*}

The X-ray spectral analysis was performed using \texttt{XSPEC v12.14.0b} \citep{arnaud96}, based on data collected post-September 4th from \textit{EP}/WXT, \textit{EP}/FXT, \nicer, \swift, and \nustar. The energy bands used for spectral fitting were 0.5–4 keV (\textit{EP}/WXT), 0.5–10 keV (\textit{EP}/FXT), 0.5–10 keV (\nicer), 0.3–10 keV (\swift), and 4–79 keV (\nustar). The \swift, \nicer, \nustar, and \textit{EP}/FXT spectra before September 20th were grouped with a minimum of 25 counts per bin, and the best fit was determined by minimizing the $\chi^2$ statistic. The \textit{EP}/FXT spectra after September 20th as well as the \textit{EP}/WXT spectra, due to the lower number of counts, were grouped with a minimum of 2 counts per bin, and the best fit was obtained by minimizing the Cash statistic \citep{Cash}. All uncertainties are reported at the 90\% confidence level for a single parameter of interest unless otherwise stated.

We began the spectral fitting using a simple absorbed power-law model (\texttt{tbabs*powerlaw} in \texttt{XSPEC}). The abundances were set to \texttt{WILM} \citep{2000Wilms} and the cross-sections were set to \texttt{VERN} \citep{1996Verner}. In most cases, all parameters including the column density $N_{\rm H}$, the photon index $\Gamma$, and the normalization are allowed to vary. However, due to the low spectral quality of the WXT observations from September 4th to 6th and the FXT observations after September 20th, $N_{\rm H}$ could not be constrained within physically meaningful ranges. Therefore, we fixed $N_{\rm H}$ at $3.1 \times 10^{22}~{\rm cm}^{-2}$, derived from the FXT observation on September 6th, which offered the best spectral quality. A simple absorbed power-law model yields acceptable fits for all spectra. The fitted parameters are summarized in Table~\ref{tab:xray_spectralfit}. We also tested the spectral fitting with the blackbody (\texttt{tbabs*bbodyrad}) or disk blackbody (\texttt{tbabs*diskbb}) model. However, neither of them provides a satisfactory fit. In addition, we performed a joint spectral fitting using the simultaneous \textit{EP}/FXT and \nustar~observations taken on September 11th. A \texttt{Constant} model was applied to correct for calibration differences between different telescopes, with \nustar/FPMA fixed at unity as the reference. Across the wide energy range of 0.5--79 keV, the spectra are still represented by the absorbed power-law model, with no evidence of iron line signatures and thermal contributions, resulting in a fit statistic of $\chi^2$/d.o.f = 1002.40/1011 and a photon index of $1.91 \pm 0.01$. On this basis, although the \texttt{powerlaw} model has already provided a statistically acceptable fit, we further tentatively test the fits and estimate the high-energy cutoff by replacing the \texttt{powerlaw} model with two models, \texttt{cutoffpl} (refer to Figure~\ref{fig:spectral_fit} for an example) and \texttt{nthcomp} \citep{Zdziarski1996, Zycki1999}, respectively.
Both models provide an equivalently-acceptable fit compared with that of the \texttt{powerlaw} model.
For the former, the cutoff energy is larger than 200\,keV. The \texttt{nthcomp} model, on the other hand, yields a stringent constraint of $T_{\rm e} > 30~{\rm keV}$, with the blackbody temperature parameter inherent in this model fixed at a typical value of 0.1\,keV. In both cases, the best-fitting column density and photon index are consistent with those obtained using the simple absorbed power-law model. In short, we conclude that the X-ray spectrum of EP J182730.0-095633 is dominated by non-thermal emission.

The unabsorbed source flux in the 0.5--10\,keV band is calculated using the \texttt{cflux} model in \texttt{XSPEC}. To determine the upper limit of the pre-outburst X-ray brightness, we stack the WXT data from September 1st to 3rd, utilizing the same source extraction and background regions as for the real detection observations. The count rate is then converted to flux using the factor derived from the first WXT detection, under the assumption that the spectral shape remained stable during the rising phase of the outburst. It is worth noting that, while this assumption may not be entirely physically justified, the main conclusions should remain unaffected. At the end of the outburst, the source was not detected in the second \swift/XRT observation. We determine the upper limit of the count rate from the 0.5--10\,keV image, and then evaluate the post-outburst flux using \textsc{webpimms} tool\footnote{\url{https://heasarc.gsfc.nasa.gov/cgi-bin/Tools/w3pimms/w3pimms.pl}} with assumptions of $N_{\rm H}=3.1\times10^{22}~{\rm cm^{-2}}$ and $\Gamma=2.0$, representing the typical spectral parameters during the outburst.

\begin{figure*}
    \centering
    \includegraphics[width=0.85\textwidth]{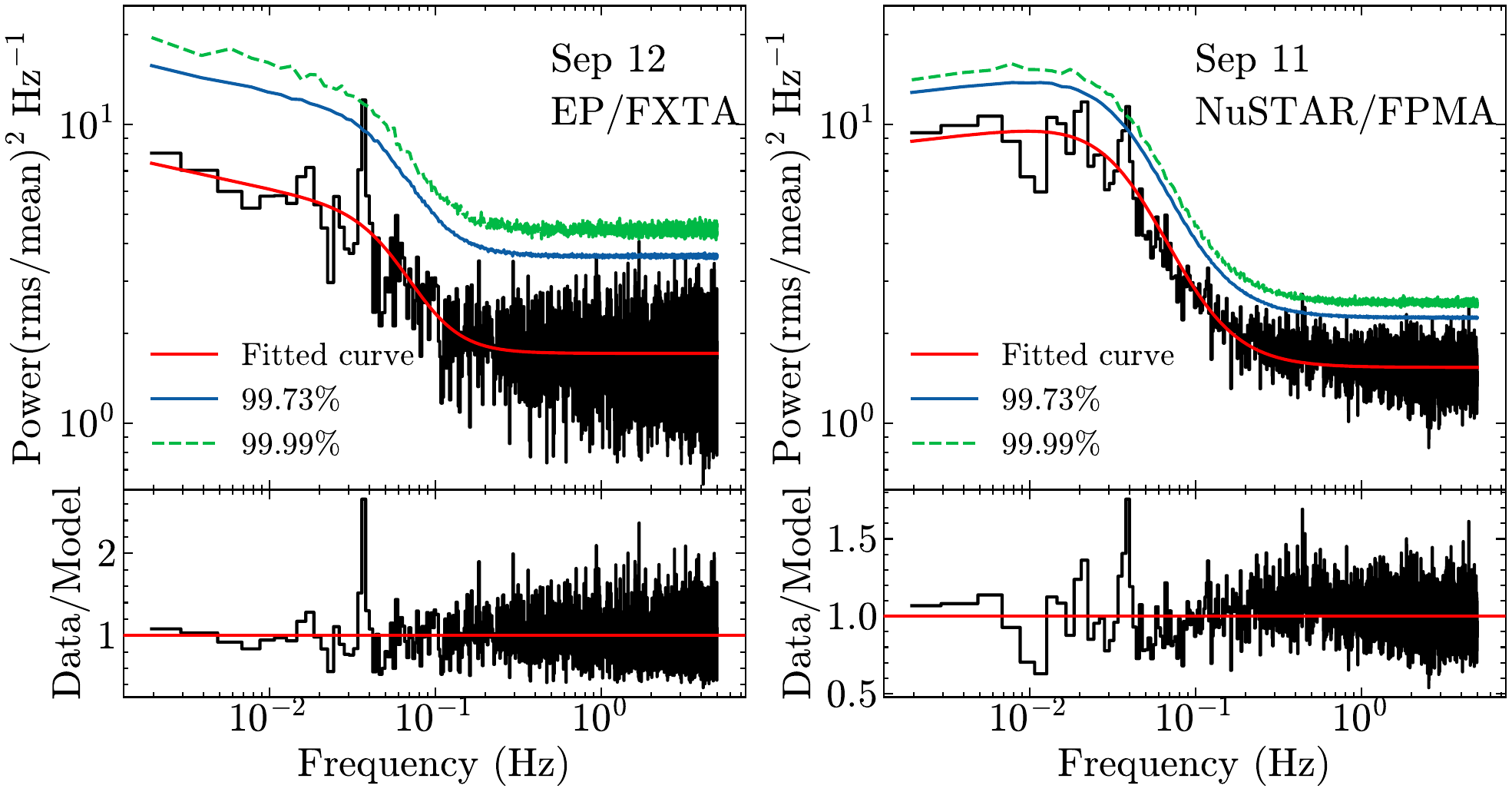}
    \caption{Power density spectra of the \textit{EP}/FXTA (left panel, ObsID 06800000091) and \nustar~data (right panel, ObsID 91001334002), with the bottom of each panel showing the power ratio between the data and best-fitting bending power-law model.}
   \label{fig:psd}
\end{figure*}

\begin{figure*}
    \centering
    \includegraphics[width=0.9\textwidth]{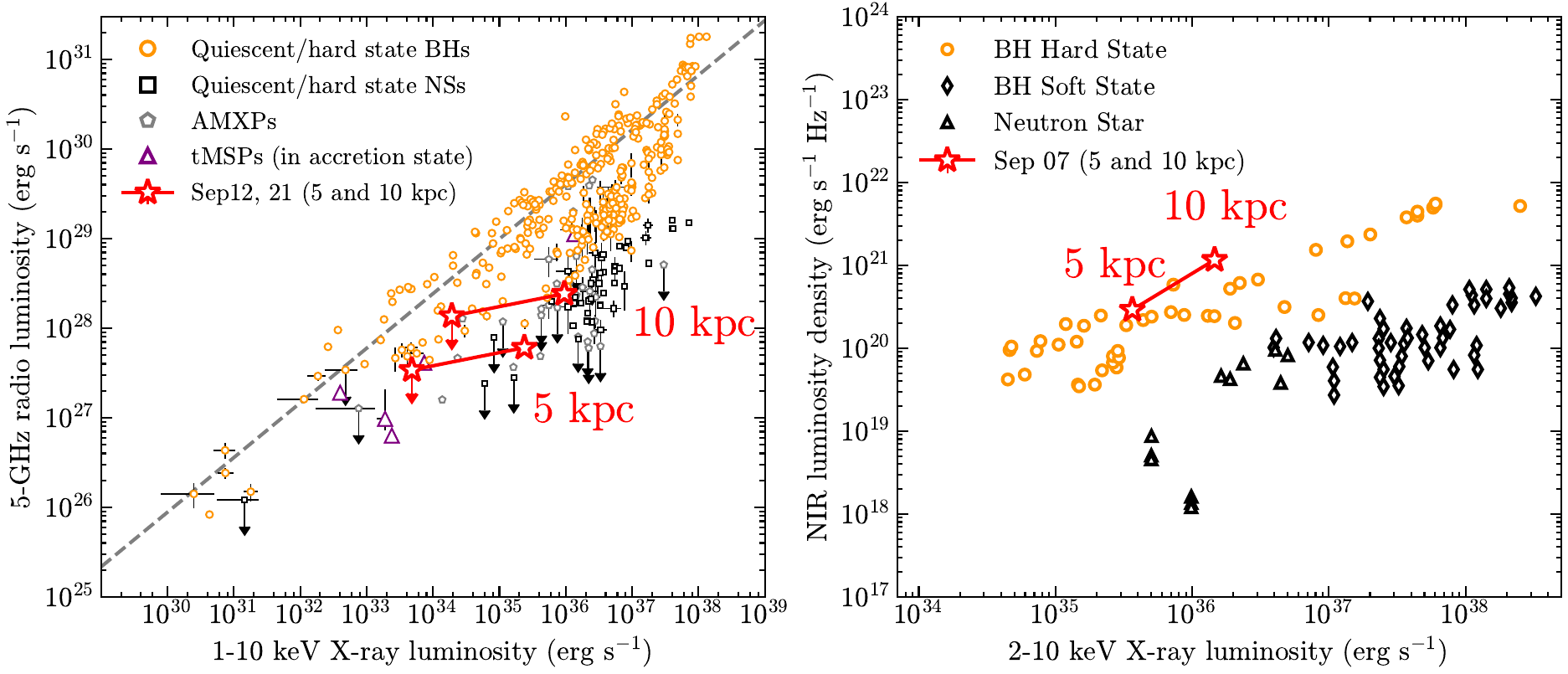}
    \caption{Quasi-simultaneous X-ray versus radio and near-infrared luminosities for LMXBs including EP J182730.0-095633. Left: the radio--X-ray ($L_{\rm R}$-$L_{\rm X}$) fundamental plane for X-ray binaries, adopted from \citet{arash_bahramian_2022_7059313}. The results of EP J182730.0-095633 are represented by red symbols, assuming various distance values. Right: the position of EP J182730.0-095633 on the near-infrared--X-ray ($L_{\rm NIR (JHK)}$-$L_{\rm X}$) correlation for X-ray binaries \citep{Russell_2006}. The data are taken on September 7th (the results barely change for those of September 8th and September 9th), and corrected for absorption using extinction values of $A_{\rm H} = 2.8\pm0.1$ mag. For brevity, only the results of the $H$ band are displayed here as a representative, with similar results for the other two bands.}
   \label{fig:lx_lr}
\end{figure*}

\subsection{X-ray temporal analysis} 
\subsubsection{Long-term X-ray behavior}
\label{sec:xray_curve}
In Figure \ref{fig:xray_evolution}, we present the long-term evolutionary trend of X-ray flux, $\Gamma$, and $N_{\rm H}$ during the outburst. The full outburst lasted approximately three weeks, during which the source's flux varied by more than three orders of magnitude. Following a rapid rise to a peak flux of approximately $5.2 \times 10^{-10}~ \text{erg}~\text{s}^{-1}~\text{cm}^{-2}$ on September 5th, the flux gradually decreased to a minimum of $\sim 2 \times 10^{-13}~\text{erg}~\text{s}^{-1}~\text{cm}^{-2}$ on September 24th. The decline phase displays a two-stage pattern. When employing a broken power-law model to fit the long-term light curve, there is an initial slow decay of $t^{-1.1}$ transitioning to a more rapid drop of $t^{-6.2}$, occurring 9.3 days after the source's discovery. Alternatively, when using a broken exponential decay model, which provides a better fit, a transition is observed at 14.1 days. Prior to this transition, the flux decreases at a rate of $e^{-\frac{t / {\rm 1~day}}{3.8}}$, which accelerates to $e^{-\frac{t / {\rm 1~day}}{1.4}}$ after the break. This dual decay pattern might correspond to different underlying physical mechanisms.

Although there are evident changes in the source's brightness, $\Gamma$ remains fairly stable, exhibiting a consistent value of $\sim2$ throughout the outburst. 
On the other hand, the column density $N_{\rm H}$ slightly decreases from $\sim3\times10^{22}~{\rm cm^{-2}}$ to $\sim2\times10^{22}~{\rm cm^{-2}}$ between September 6th and September 17th, suggesting that the origin of X-ray absorption is intrinsic to the source. 
The observed excess above the Galactic HI column density in the direction of the source, which is around $1.13\times10^{22}~{\rm cm^{-2}}$ as determined using the HI4PI Survey map \citep{HI4PI} accessed through the HEASARC online tool \footnote{\url{https://heasarc.gsfc.nasa.gov/cgi-bin/Tools/w3nh/w3nh.pl}}, also supports this interpretation.
We also calculated the evolutionary trend of the spectral index while keeping the column density fixed at the typical value of $3.1\times10^{22}~{\rm cm^{-2}}$, yielding results consistent with those presented above within measurement uncertainties (see Figure \ref{fig:gamma_evolution_with_nh_fixed} for an illustration).

\subsubsection{Short time-scale variability}
\label{sec:periodicity}
We conducted a search for thermonuclear X-ray bursts and pulsations in all available X-ray data, as these phenomena serve as strong evidence for identifying the source in a neutron star (NS) system. However, no thermonuclear X-ray bursts were identified in the light curves. 
Additionally, no significant pulsations were detected in the power spectra, with 3$\sigma$ upper limits on the flux pulsed fraction of 6\%--10\% for the second \nicer\ observation, 8\%--13\% for the \nustar\ observation, and 6\%--13\% for the \textit{EP}/FXT observations across the frequency range 30--1000\,Hz (for more details, see Appendix\,\ref{sec:xray_periodic_search}).
Although the light curves do not show thermonuclear X-ray bursts or pulsations, they present strong short-time variations. 
We conducted a power spectral density (PSD) analysis using a timing resolution of 0.1\,s, and a segment length of 256~s or 512~s. For most observations a segment length of 512~s is adopted, while for observations with relatively short GTIs, a 256~s segment length is chosen for a higher signal-to-noise ratio in resulting PSD.
The PSD is normalized in units of $\rm rms^{2}$ per Hz \citep{Belloni_1990}. Figure~\ref{fig:psd} displays the PSDs for the observations taken by \textit{EP}/FXTA and \nustar/FPMA. The continuum component of both PSDs exhibits increased noise towards lower frequencies (red noise) and can be well modeled by a bending power-law function with a break frequency of around 0.03–0.06\,Hz. Additionally, there is a prominent, narrow peak around 0.038\,Hz, indicating the presence of a quasi-periodic oscillation (QPO) with an RMS amplitude of $\sim$20\%. 
To assess the significance of the QPO, we simulated one hundred thousand light curves, based on the method of \citet{TimmerKonig1995} from the best-fit bending power-law PSD.
We then calculated the PSDs of these light curves, and determined the distribution of the variability power at each Fourier frequency. 
The significance of the QPO greatly exceeds the $3\sigma$ confidence threshold, with the quality factor $q\equiv\nu_0/(2\Delta{\nu})$ estimated to be $\gtrsim8$. The QPO is also present in the FXT module B (FXTB) data, collected in the Partial Window (PW) mode, with a centroid frequency, significance, and {\it q}-factor that roughly match those found in the FXTA data. The QPO feature is evident in other FXT observations before September 20th, as shown in Figure~\ref{fig:FXT_PSD}, with the centroid frequency remaining around 0.04 Hz despite the X-ray flux varying by more than two orders of magnitude. After September 20th, no clear QPO signal was observed, probably because of the very low count rate of the source. In addition, a similar QPO is also detected in the \nicer~data, although with lower significance of about 2 $\sigma$ (Figure~\ref{fig:FXT_PSD}).

\subsection{Optical and infrared detection}
\label{sec:optical_detection}

\textit{SVOM}/VT rapidly responded to observe the field of EP J182730.0-095633 on September 6th, yet no uncatalogued objects were detected within the \textit{Swift}/XRT error circle. The $5\sigma$ upper limits for the $VT_{\rm R}$ (650 - 1000 nm) and $VT_{\rm B}$ (400 - 650 nm) bands are $21.4$ mag and $22.9$ mag, respectively. From September 7th to 9th, this field was extensively observed by several instruments, including GROND, NOT, Mephisto, TRT and ALT100C. The observations of Mephisto, TRT and ALT100C did not uncover any new or flaring objects around the X-ray position, but an uncatalogued source was detected by GROND in the near-infrared (NIR) $J$, $H$ and $K_{s}$ bands in all three epochs (see panels (e) and (f) in Figure~\ref{fig:ep_detection_image}), with consistent magnitudes of $J\sim18$ mag, $H\sim16.8$ mag and $K_{s}\sim16.4$ mag, all in the AB system (see Table \ref{tab:EP J182730.0-095633_opt_infrared_followups}). 
The position of the NIR signal is R.A. = $18^{\rm h}27^{\rm m}30.09^{\rm s} \pm 0.01^{\rm s}$ and Decl. = $-9^{\circ}56'39.6'' \pm 0.2''$.
It is worth noting that the signal was also marginally detected by NOT in the $r$ band on September 7th (with a magnitude of $23.87\pm0.35$ mag; \citealt{2024ATel16807....1F}), as well as by GROND in both $r^{\prime}$ and $i^{\prime}$ bands on September 8th (with magnitudes of $23.86\pm0.30$ mag and $21.77\pm0.10$ mag, respectively). Interestingly, this source was no longer detected in the following GROND observations carried out on September 11th and thereafter, clearly indicating a fading behavior. Additionally, at the end of the X-ray outburst, \textit{SVOM}/VT and TNT carried out two additional observations for this field on September 28th and October 11th, respectively, yet no sources were detected. The long-term evolution of the apparent magnitudes during the X-ray outburst is shown in the bottom panel of Figure \ref{fig:xray_evolution}.

The faint optical and near-infrared detection is likely due to high extinction, given the low Galactic latitude of EP J182730.0-095633. Specifically, in the source direction the Galactic foreground reddening in the $V$ band is very severe, with $A_{\rm V}\sim15.4-15.8$ mag \citep{SloanDigitalSkySurvey} accessed through dust extinction tools on NED\footnote{\url{https://ned.ipac.caltech.edu/extinction\_calculator}} and IRAS\footnote{\url{https://irsa.ipac.caltech.edu/applications/DUST/}}. The reddening for the GROND $J$ and $H$ band are estimated to be $4.4\pm0.1$ mag and $2.8\pm0.1$ mag, respectively, by applying the relationship between $A_{\lambda}$ and $A_{\rm V}$ taken from \citet{Cardelli_1989}. We take these values for subsequent NIR luminosity calculations (see Section~\ref{sec:discussion}). The results of GROND and \textit{EP}/FXT observations during when the source was detected by GROND from September 7th to 9th were taken for the investigation of the $L_{\rm X}$-$L_{\rm R}$ correlation, as shown in the right panel of Figure \ref{fig:lx_lr}.

\subsection{Radio detection}
\label{sec:radio_detection}
The first \textit{ATCA} observation on September 12th identified a radio source at the X-ray position (Figure~\ref{fig:ep_detection_image}). At 5.5\,GHz, the source position was R.A. = $18^{\rm h}27^{\rm m}30.16^{\rm s} \pm 0.39^{\rm s}$ and Decl. = $-9^{\circ}56'36.0'' \pm 3.51''$, while at 9 GHz, it was located at R.A. = $18^{\rm h}27^{\rm m}30.07^{\rm s} \pm 0.21^{\rm s}$ and Decl. = $-9^{\circ}56'40.4'' \pm 2.09''$. The radio locations are $5.4\pm4.1''$ and $1.0\pm3.0''$ away from the \textit{Swift} X-ray position respectively. The peak flux densities at 5.5\,GHz and 9\,GHz were measured to be $101.2\pm5.4$\,µJy/beam and $126.3\pm5.2$\,µJy/beam, respectively. The radio spectral index ($\alpha$) is $0.45\pm0.14$, showing an inverted spectrum. During a subsequent \textit{ATCA} observation, conducted on September 21st, no significant radio emission was detected, yielding 3$\sigma$ upper limits of 57\,$\mu$Jy/beam at 5.5\,GHz and 60\,$\mu$Jy/beam at 9\,GHz. A slight positional offset ($\sim4.5''$) was observed between the centroids at the two frequencies in the first \textit{ATCA} observation, with a significance level of $3~\sigma$. 
It remains unclear whether this discrepancy is due to intrinsic physical processes or systematic errors from instrumental effects, as the source was undetectable in later observations. 
The results of \textit{EP}/FXT and \textit{ATCA} observations on September 12th and 21st were taken for the investigation of the $L_{\rm X}$-$L_{\rm R}$ correlation, as shown in the left panel of Figure \ref{fig:lx_lr}.

In the first \textit{MeerKAT} epoch we obtained a detection at S-band (3.06\,GHz) in which the source located at R.A. = $18^{\rm h}27^{\rm m}30.00^{\rm s} \pm 0.009^{\rm s}$ and Decl. = $-9^{\circ}56'39.75'' \pm 0.13''$, with a flux density of $33 \pm 4$ $\mu$Jy. While at L-band (1.28\,GHz) we have a non-detection, with a $3\sigma$ upper limit of 69\,$\mu$Jy/beam. 
The source position obtained by \textit{MeerKAT} is consistent with those provided by \textit{Swift}/XRT and \textit{ATCA} at 9\,GHz.
The spectral shape of \textit{MeerKAT} is also consistent with the positive spectral index obtained in the first \textit{ATCA} epoch.
In our second \textit{MeerKAT} epoch we obtain a detection at L-band at $58 \pm 12$ $\mu$Jy, and a non-detection at S-band with a 36\,$\mu$Jy/beam $3\sigma$ upper limit.
Finally, we obtained no-detections at both L- and S-band in our third \textit{MeerKAT} epoch, with 105\,$\mu$Jy/beam and 36\,$\mu$Jy/beam upper limits, respectively. The results of \textit{ATCA} and \textit{MeerKAT} are summarized in Table \ref{tab:EP J182730.0-095633_radio_followups}.


\section{Discussion}
\label{sec:discussion}

\subsection{Nature of the transient}
EP J182730.0-095633 (EP240904a) is a new X-ray transient discovered by \textit{EP}/WXT on September 4th, 2024. Follow-up observations were performed with multiple X-ray telescopes and multi-wavelength facilities. In this paper, we report on the spectral and temporal properties of this source during the outburst, aiming to uncover its physical nature.

At the corresponding location of EP J182730.0-095633, a faint NIR transient source was detected by GROND and NOT. The NIR flux showed a rapid decay and became undetectable within a few days, accompanied by a simultaneous decline in the X-ray flux. Given its spatial and temporal coincidence, we suggest that the transient source detected by GROND and NOT is very likely to be associated with EP J182730.0-095633, thus representing a promising candidate for the NIR counterpart. We note that the non-detection of the signal in optical bands can be attributed to the large Galactic foreground reddening of $A_{\rm V}\approx15.4-15.8$ mag along the line-of-sight. 

Due to the faintness and rapid diminishing of the potential counterpart, no optical/NIR spectrum was obtained, and thus no redshift measurement could be made.
The absence of a redshift measurement complicates the task of determining the source's nature. However, it is largely unlikely that this transient is either a Gamma-Ray Burst (GRB) or a stellar flare due to the differing characteristics. GRBs generally exhibit a much shorter timescale for their rise and decay, with the former typically occurring within a few seconds and the latter lasting a few days \citep{Zhang_grb}, whereas stellar flares are characterized by X-ray spectra primarily described by one or several optically thin thermal plasma components \citep{Benz_2010}. 

The tidal disruption event (TDE) may be a potential scenario, given that some of the observational properties of EP J182730.0-095633, such as the non-thermal X-ray spectrum, two-stage decaying trend as well as the radio and infrared detections, are broadly consistent with a rare class of TDEs dominated by relativistic jet \citep[e.g.][]{Burrows2011, Bloom2011, Pasham2015}. However, the infrared-to-X-ray flux ratio appears to be significantly higher compared to that observed in jetted TDEs \citep[e.g.][]{Pasham2015, Levan2016}. Meanwhile, we do not observe a plateau phase or strong intra-day variability in X-rays which instead are common features in those jet-dominated systems. Therefore, the possibility of EP J182730.0-095633 being a jetted-TDE seems less unlikely. Moreover, it is interesting to point out that, despite numerous differences, EP J182730.0-095633 has exhibited a PSD profile similar to a rare white dwarf (WD) - terrestrial-icy planet TDE candidate IGR J17361-4441 \citep{Bozzo_2011, J17361_TDE} discovered by the IBIS/ISGRI telescope \citep{Ubertini2003} onboard \textit{INTEGRAL} on 2011 August 11, which shows a notable QPO signal with a centroid frequency of  $\sim100$ mHz \citep{Bozzo_IGRJ17361}.

In contrast, EP J182730.0-095633 resembles XRBs in a variety of aspects regarding the source position, radio emission, X-ray variability, spectral properties, and multi-wavelength relationships. Firstly, the position of EP J182730.0-095633 is on the Galactic plane, a region densely populated by X-ray binaries \citep[e.g.][]{Grimm_2002, Bahramian_LMXB}. Secondly, the X-ray flux of EP J182730.0-095633 generally follows a fast-rise-exponential-decay behavior (FRED), which agrees largely with the canonical outburst behavior of XRBs \citep{Chen_FRED, Remillard_review}. The multifold pattern during the decaying phase can also be interpreted within the context of XRB outbursts \citep[e.g.][]{King_Ritter_1998, Dubus2001, Weng2015}. Thirdly, as shown in Figure~\ref{fig:lx_lr}, EP J182730.0-095633's positions on the $L_{\rm X}$-$L_{\rm R}$ and $L_{\rm X}$-$L_{\rm NIR}$ diagrams are in good agreement with an XRB origin, assuming a distance of several kpc. Additionally, the radio detection, combined with an inverted radio spectrum obtained by \textit{ATCA} and \textit{MeerKAT}, may suggest the presence of a compact jet, which is commonly observed in the hard states of XRBs \citep[see][and references therein]{Belloni_jet}. This possibility is further supported by the observed non-thermal power-law spectrum, a characteristic feature in the hard states. The spectral shape variation in the radio band during the late outburst stage implies that the compact jet transitioned from optically thick to optically thin state (see Section \ref{sec:radio_detection}), a phenomenon commonly observed during the decay phase of XRB outbursts \citep[e.g.][]{Corbel2013_corr, 2018Trigo}. More importantly, the QPO signal, which is a prevalent feature in XRBs \citep{Ingram_Motta_2019}, is significantly detected during a long period of the outburst. 

\subsection{Properties of the accreting system}

If EP J182730.0-095633 is a Galactic XRB, the peak X-ray luminosity ($L_{2-10~{\rm keV}}$) is estimated to be $\lesssim 10^{36}~{\rm erg~s^{-1}}$, assuming a distance of several kpc. According to \citet{Wijnands_2006}, it can then be classified as a borderline faint to very faint XRB (VFXB). These sources were recently and extensively discovered in the Galactic Center, Galactic Plane, and Galactic Bulge by {\it ASCA}, {\it Swift}, {\it Chandra} and {\it XMM-Newton} surveys \citep[e.g.][for a detailed review, one may refer to \citet{Bahramian_LMXB}]{Hands2004, Muno2005a, Muno2005b,Sakano2005, Degenaar2009}. Their outbursts typically last for several weeks to months \citep[e.g.][]{Degenaar2009, Heinke2015, Weng2015}.

Identifying the nature of the central compact object (NS or BH) is somehow challenging for EP J182730.0-095633, due primarily to the lack of definitive evidence supporting either scenario (e.g. coherent pulsations, thermonuclear bursts or dynamical mass estimation). 
Its locations on the radio vs. X-ray luminosity fundamental plane (see left panel of Figure \ref{fig:lx_lr}) fall onto the overlapping regions of BHs and NSs, yielding inconclusive argument.
Nevertheless, we are inclined to argue that EP J182730.0-095633 is more likely to be a BHXRB for the following reasons. Firstly, the X-ray emission throughout the outburst is consistently described by a non-thermal power-law spectrum with a photon index maintaining at $\sim2$. This `stable' behavior largely precludes the source being a NS accreting system as the spectrum of the latter often exhibits a softening trend as the luminosity decreases below $10^{36}~{\rm erg~s^{-1}}$ due to the presence and growing prominence of the NS surface emission \citep{Wijnands2015}. This dichotomy in the spectral shape evolution has been increasingly studied in recent years \citep[e.g.][]{Plotkin2013, Wijnands2015, Qiao2020}, and developed to be a diagnostic tool for identifying BH (candidates) \citep[e.g.][]{Stoop2021}.
Secondly, the relative strength of the X-ray and NIR emissions (Figure~\ref{fig:lx_lr}) closely aligns with those typically observed in BH binaries in the hard state \citep{Russell_2006}. 
Thirdly, the joint spectral fitting of EP/FXT+\nustar ~data indicates a relatively high electron temperature of $T_{\rm e}>30$ keV (see Section \ref{sec:spectral}), which is more prone to exist in BH systems \citep[e.g.][]{Burke2017, Banerjee2020}.
Furthermore, even though mHz QPOs are detected in a number of NS systems and are usually attributed to the marginally stable nuclear burning on the NS surface \citep[e.g.][]{Revnivtsev_mhzQPO_2001, Strohmayer_mhzQPO_2011, Mancuso2019,Lyu2020, Mancuso2023, Tse2023}, the only two VFXBs with QPO detection, Swift J1357.2-933 \citep{SwiftJ1357.2-0933_QPO,Beri_1357_QPO} and XTE J1118+480 \citep{XTEJ1118_QPO,Wood_XTE_J1118+480_QPO}, have been confirmed to host BH accretors.
Future observations, particularly dynamical mass measurements, are essential to verify this scenario and unveil the mysterious nature of EP J182730.0-095633.

Regarding the companion star, its nature remains largely inconclusive due to the current lack of spectroscopic information. However, we can tentatively explore this issue based on the quasi-quiescent NIR luminosity, assuming that it is contributed mainly by the companion. Specifically, considering the upper limit of $m_{\rm J} > 19.2~{\rm mag}$ given by GROND observations on September 18th (near the end of the X-ray outburst), the corresponding absolute magnitude upper limit is $M_{\rm J}>4.7~{\rm mag}$ assuming a distance of 10 kpc and a Galactic foreground dust extinction of $A_{\rm J}$ = 4.4\,mag. According to typical stellar parameters \citep{Pecaut2012, Pecaut2013}, the companion star is likely a late type K/M star, indicating that this source is a potential low-mass XRB.

The X-ray outburst of EP J182730.0-095633 resembles the `hard-only' outbursts widely observed in X-ray binaries \citep[e.g.][]{Yan2015, watchdog, Alabarta2021}, which are notably characterized by the absence of state transitions.
Its X-ray spectra show no iron ${\rm K\alpha}$ line or thermal component, indicating that the standard Shakura-Sunyaev disk \citep[SSD,][]{SSD} is either absent or truncated at a large radius. 
The accretion close to the compact object is likely to operate in the form of an optically thin, geometrically thick advection-dominated accretion flow \citep[ADAF,][]{Ichimaru1977,Narayan1994, Narayan95a, Narayan95b, Yuan_ADAF}.
The observed X-ray spectral index remains relatively constant throughout the entire outburst, suggesting that the geometry of the accretion flow is relatively stable during the flaring activity. In such a scenario, the mHz QPOs, potentially generated by Lense-Thirring precession or quasi-periodic obscurations \citep[as discussed in][]{Beri_1357_QPO}, could exhibit stable frequencies, as observed.

We also note that there is another possibility that EP J182730.0-095633 is an extragalactic XRB system. If the correlation between X-ray and radio emissions is expected to follow the $L_{\rm X}$-$L_{\rm R}$ relation typical of XRBs, its estimated distance would be within $\sim 10^2$\,kpc. However, there are no nearby satellite galaxies in the direction of EP J182730.0-095633 \citep{Drlica-Wagner2020}, making this scenario less likely. 

\subsection{mHz QPO signal}
\label{sec:mhzqpo}
Despite of the ubiquitous nature of QPOs in BHXRBs across various spectral states, the detection of the mHz QPO remains relatively rare \citep{Ingram_Motta_2019}, with only a few cases reported in the past few years, e.g. MAXI J1820+070 \citep{2025Li_MAXJ1820}, H 1743--322 \citep{Altamirano2012} and MAXI J1348--630 \citep{2024WangXinLei}.
The centroid frequency of the mHz QPO, falls within the range of type-C low-frequency QPOs (LFQPO), the most common type of QPO in X-ray binaries \citep[e.g.][]{Motta2011, Buisson2019}.
To tentatively probe the association of this $\sim40$ mHz QPO signal with the commonly-observed type-C QPOs, we calculate the fractional-rms spectrum of QPO using the data of \textit{EP-FXT} and \textit{NuSTAR} obtained on September 11th 2024, as shown in the left panel of Figure \ref{fig:qpo_discussion}. 
Regardless of a large uncertainty, the rms-energy spectrum exhibits an enhanced fractional variability with the increasing photon energy below $\sim20$ keV, then tends to flatten toward higher energies.
This is generally consistent with that observed in type-C LFQPOs of XRBs \citep[e.g.][]{Casella2004, Rodriguez2004, Yadav2016, 2024YangZX}.
We further calculate the relation between the QPO frequency and the low-frequency PSD break for QPO detections with significance exceeding $3\sigma$, as presented in the right panel of Figure \ref{fig:qpo_discussion}.
The measurement of the EP transient is denoted in blue circles and the canonical W-K relation built for type-C LFQPOs \citep{1999WijnandsvanderKlis} is signified in black dashed line.
Notably, the mHz QPO of EP J182730.0-095633 exhibits a much lower centroid frequency than predicted by the empirical W-K relation, also the QPO frequency remains quite stable despite of the variations in the PSD break.
This makes it much less likely to be a typical type-C QPO.
In this context, the QPO detected in EP J182730.0-095633 may represent a new class of LFQPOs in BHXRBs.
It is also interesting to point out that the relative position of break and QPO looks similar to the low-frequency QPOs found in some Ultra-Luminous X-ray sources \citep[ULXs, e.g.][]{Dewangan2006,Atapin2019,ElByad2025}.
We note that a detailed investigation on the origin of this $\sim40$ mHz QPO is beyond the scope of this paper and will be deferred to future work.

\vspace{0.5cm}

\section{Conclusions}
\label{sec:conclusions}
In this \textit{Letter}, we report on the discovery and multi-wavelength follow-up observations of an intriguing X-ray transient EP J182730.0-095633 (EP240904a). 
This transient was first detected by \textit{EP}/WXT on September 4th, 2024, with its location on the Galactic plane ($l=21.3634\deg$, $b=0.7244\deg$). The source underwent a faint ($F_{\rm peak,~0.5-10~keV}\approx5.2 \times 10^{-10}~{\rm erg~s^{-1}~cm^{-2}}$), short-lived ($\sim20$ days) outburst.
The X-ray emission is of non-thermal origin, with the spectrum well described by a power-law model with a photon index ($\Gamma$) maintaining at $\sim2$. 
A relatively high electron temperature ($T_{\rm e}>30~{\rm keV}$) is inferred from the joint fit of \textit{EP}/FXT+\textit{NuSTAR} spectra using the \texttt{nthcomp} model.
No thermonuclear bursts or significant coherent pulsations are detected in X-ray data, while a long-lasting millihertz QPO signal with a centroid frequency of $\sim0.04$\,Hz is significantly detected by various X-ray instruments. A radio source displaying an inverted spectrum ($\alpha = 0.45\pm0.14$), as well as a potential, transient NIR counterpart, are detected at the X-ray position. 

Based on the source position, X-ray spectral and timing properties (stable evolution of $\Gamma$, high electron temperature, and the detection of mHz QPO), the radio spectral shape, as well as the multi-wavelength properties (the $L_{\rm X}$-$L_{\rm R}$ and $L_{\rm X}$-$L_{\rm NIR}$ correlations and the quasi-quiescent NIR brightness), we conclude that EP J182730.0-095633 is very likely to be a new black hole low-mass X-ray binary captured during a faint outburst without state transitions.
The discovery of EP J182730.0-095633 showcases \textit{EP}'s potential to find transient BHs exhibiting faint X-ray outbursts, which represent a potentially large population yet are mostly elusive to the other X-ray wide-field monitoring instruments. 
More of such systems are expected to be discovered in the \textit{EP} era, which may advance our understandings of the underlying physical mechanism of faint-level outbursts, and hopefully shed light on the ``missing" puzzle of black holes in our galaxy.

\facilities{\emph{Einstein Probe} \citep{Yuan22}, \swift\ \citep{Gehrels_Swift}, \nicer\ \citep{Gendreau_2016}, \nustar\ \citep{Harrison_nustar}, \hxmt\ \citep{Insight_HXMT_Zhang}, \textit{SVOM} \citep{2016SVOM}, \textit{GROND} \citep{GROND2008}, \textit{Mephisto} \citep{2020Mephisto}, \textit{TNT} \citep{Huang2012TNT}, \textit{ATCA}, \textit{MeerKAT} \citep{Camilo2018, Jonas2018}.}

\software{WXTDAS v1.0, fxtdas v1.10, HEASOFT v6.33.2 \citep{heasoft}, Matplotlib v3.9 \citep{hunter07}, NICERDAS v12, NuSTARDAS v2.1.4, XSPEC v12.14.0h \citep{arnaud96}, PRESTO \citep{Ransom_2002}, Source Extractor v2.28.0 \citep{1996SourceExtractor}, IRAF v2.16 \citep{1986IRAF}, CASA \citep{2022CASA}.}

\section*{Acknowledgments}
\label{sec:acknowledgement}
We thank the anonymous referee for comments that help improve the paper. This work is based on data obtained with \ep, a space mission supported by Strategic Priority Program on Space Science of Chinese Academy of Sciences, in collaboration with ESA, MPE and CNES.
This work is supported by the National Natural Science Foundation of China (Grant Nos. 12203071, 12333004, 12122306, 12173048, 12473047, 12473016, 12373049, 12433007, 12361131579), the National Key Research and Development Program of China (Grant Nos. 2023YFA1607903, 2024YFA1611603, 2024YFA1611604), the National SKA Program of China (2022SKA0130100),  the Tianshan Talent Training Program (2023TSYCCX0101), and the Strategic Priority Research Program of the Chinese Academy of Sciences (Grant No. XDB0550200).
FCZ, AM and NR are supported by the European Research Council (ERC) via the consolidator grant ``MAGNESIA'' (No. 817661; PI: Rea), the proof of concept grant ``DeepSpacePULSE'' (No. 101189496;PI: Rea), the Spanish grant PID2023-153099NA-I00 (PI: Coti Zelati), the Catalan grant SGR2021-01269 (PI: Graber/Rea), and the program Unidad de Excelencia Mar\'ia de Maeztu CEX2020-001058-M. FCZ also acknowledges support from a Ramon y Cajal fellowship (grant agreement RYC2021-030888-I). RS acknowledges support from the INAF grant number 1.05.23.04.04. JPUF is supported by the Independent Research Fund Denmark (DFF--4090-00079) and thanks the Carlsberg Foundation for support. The Cosmic Dawn Center (DAWN) is funded by the Danish National Research Foundation under grant DNRF140. AB acknowledges support from the UK Space Agency. P.G.J.~is supported  by the European Union (ERC, Starstruck, 101095973). Views and opinions expressed are however those of the author(s) only and do not necessarily reflect those of the European Union or the European Research Council Executive Agency. Neither the European Union nor the granting authority can be held responsible for them.

This work is also based on data obtained by: the Neil Gehrels \swift\ Observatory (a NASA/UK/ASI mission) supplied by the UK \swift\ Science Data Centre at the University of Leicester; \nicer, a 0.2--12\,keV X-ray telescope operating on the International Space Station, funded by NASA; the \nustar\ mission, a project led by the California Institute of Technology, managed by the Jet Propulsion Laboratory, and funded by NASA. 
We thank the \nicer\ PI, Keith Gendreau, for approving our Target of Opportunity (ToO) request and the operation team for executing the observations; we thank the \swift\ deputy project scientist, Brad Cenko, and the \swift\ duty scientists and science planners, for making the \swift\ ToO observations possible; we also thank the \nustar\ PI, Fiona Harrison, for approving our Director's Discretionary Time request, and the \nustar\ science operation center for carrying out the observation.
The Australia Telescope Compact Array is part of the Australia Telescope National Facility (grid.421683.a) which is funded by the Australian Government for operation as a National Facility managed by CSIRO.We acknowledge the Gomeroi people as the traditional owners of the Observatory site. The MeerKAT telescope is operated by the South African Radio Astronomy Observatory, which is a facility of the National Research Foundation, an agency of the Department of Science and Innovation. This work has made use of the “MPIfR S-band receiver system” designed, constructed and maintained by funding of the MPI für Radioastronomy and the Max-Planck-Society.

We acknowledge the data resources and technical support provided by the China National Astronomical Data Center, the Astronomical Science Data Center of the Chinese Academy of Sciences, and the Chinese Virtual Observatory. 
The storage and processing of the EP data are supported by Alibaba Cloud through its Creation Beyond Cloud Program.
Part of the funding for GROND (both hardware as well as personnel) was generously granted from the Leibniz-Prize to Prof. G. Hasinger (DFG grant HA 1850/28-1)

\appendix

\section{Multi-waveband Data Reduction}
\label{sec:app:data}

\subsection{X-rays}
\label{sec:detailed_xraydatareduction}

\subsubsection{\textit{Einstein Probe}\,}\label{sec:ep}

EP J182730.0-095633 was first detected in an observation starting on September 4th, 2024, at 10:12:56 UT and lasting approximately 8.8 ks, on the CMOS detector No.39 of WXT (ObsID: 06800000067). Previous individual observations covering three days before the detection, as well as the stacked data, did not reveal any notable source signals. The WXT images from September 3rd to 4th (first detection) are shown in panels (a)--(b) of Figure \ref{fig:ep_detection_image}. It is worth noting that for the observation on September 4th, the emission of the source is rather stable without short-term flaring activity. Following the discovery, an extensive monitoring campaign was conducted by \textit{EP}/FXT from September 6th to 24th, with a cadence of twice per day during the first week and $\sim$ once per day thereafter. A total of 22 FXT observations were conducted and utilized for analysis. We note that the source was too faint for detection on September 23rd (ObsID: 06800000118), due mainly to a significantly shorter exposure time of approximately $\sim1000~{\rm s}$. The FXT was operated with module A (FXTA) in Timing Mode (TM) and module B (FXTB) in Partial Window (PW) mode, except during the first observation, where both FXTA and FXTB were in Full Frame (FF) mode. 
The first FXT observation provides a more precise localization of the source, with coordinates of R.A. (J2000.0) = $18^{\rm h}27^{\rm m}30.0^{\rm s}$ and Decl. (J2000.0) = $-9^{\circ}56'33''$ (with an uncertainty of $10''$ at the 90\% confidence level).

For \textit{EP}/WXT, we take observations from September 1st to 6th for data analysis. Data reduction was performed using \texttt{wxtpipeline}, the standard pipeline of the WXT Data Analysis Software (\texttt{WXTDAS}, Liu et al. in preparation), and the calibration database (CALDB) developed by the EP Science Center (EPSC). The CALDB is built based on the results of the on-ground calibration experiments \citep{2025ChengEPWXT}, with the procedure successfully implemented on a prototype of the WXT instrument \citep{Cheng2024_LEIA_calibration}.
\texttt{wxtpipeline} performs calibrations to the original event data, including the coordinate transformation, flagging bad and hot pixels, and computing the pulse height amplitude invariant (PI) values, then generates the clean event list by screening the events based on the default criteria (BR\_EARTH$>=25$~\&ELV$>=10$~\&SAA$==0$~\&ANG\_DIST$<0.1$). The sky image and exposure map are extracted from the clean event list. A point-source detection is then implemented on the sky image. Sources of high significance \citep[greater than $5 \sigma$, calculated using the Li-Ma method,][]{LiMa1983} are cataloged, creating a detection list. For each identified source, several outputs are produced, including the light curve and spectra, the ancillary response file, the response matrix, as well as the background light curve and spectra. The light curves and spectra are extracted using a circular region centered on the source, with a radius of $9.1'$. For the background, an annular region centered on the source is employed, with inner and outer radii of $18.2'$ and $36.4'$, respectively. The \texttt{wxtmerge} tool is utilized to stack multiple WXT observations.

Data reduction for FXT was performed using \texttt{fxtchain}, the standard data analysis pipeline within the FXT Data Analysis Software \citep[\texttt{FXTDAS,v1.10},][]{2025ZhaoFXTDAS} developed by EPSC. This tool includes several procedures, including: particle event identification, pulse-invariant value calculation, bad and hot pixel flagging, and the selection of good time intervals (GTIs) using housekeeping data. The source was bright in the first observation carried out in FF mode, leading to a non-negligible pile-up effect. To mitigate this effect the source photons were extracted using an annulus region centered on the target with an inner radius of $15''$ and outer radius of $80''$. The background photons were taken from a nearby source-free circular region with a radius of $150''$. For the PW imaging mode, the radius of the source and background circular regions were both set to $60''$. For the TM mode, the source and background photons were extracted using two rectangular regions centered on the source and away from the source, respectively, both with a size of $3'\times1'$. 
The Solar System barycentric correction for photon arrival times was applied using the \texttt{fxtbary} tool. The TM data were excluded from the spectral analysis because of a lower spectral quality compared to those taken in the imaging mode.

\subsubsection{NICER}\label{sec:nicer}
Following the discovery of the source, \textit{NICER} made several observations beginning on September 7th (PI: Huaqing Cheng). However, from September 23rd onwards (ObsID 7204340113), the source's flux fell below $10^{-12}~\text{erg}~\text{s}^{-1}~\text{cm}^{-2}$, resulting in the NICER/XTI spectra with a considerable background noise. Consequently, these observations were excluded from the data analysis. For the observations before September 23rd, clean event files were produced using \texttt{nicerl2} with specific settings: \texttt{underonly\_range="0-200"} and \texttt{overonly\_range="0-2"}. Focal Plane Modules (FPMs) 14 and 34 were not included in the analysis as they often show elevated detector noise. Light curves and spectra were extracted using \texttt{nicerl3-lc} and \texttt{nicerl3-spec}, respectively, employing the \textsc{scorpeon} background model \footnote{\url{https://heasarc.gsfc.nasa.gov/docs/nicer/analysis\_threads/scorpeon-overview/}}. This rigorous selection process resulted in only 4 observations (ObsIDs: 7204880101, 7204880102, 7204880103, and 7204880105) with clean events. Photon arrival times were corrected to the Solar System barycenter using the \texttt{barycorr} tool.

\subsubsection{Swift}
\label{sec:swift}
We conducted two target-of-opportunity (ToO) observations using the X-ray telescope (XRT) onboard the Neil Gehrels Swift Observatory (\textit{Swift}) in the Photon Counting (PC) mode, one on September 8th, 2024 and another on November 1st, 2024 (PI: Huaqing Cheng). 
The source was significantly detected in the first \swift~ observation, while no signals were found in the second epoch. The XRT data were processed using \texttt{xrtpipeline}, and the spectra were extracted from the level-2 cleaned data products using \texttt{xselect}. The source photons were initially extracted from a circular region centered on the target with a radius of \(47''\), while the background photons were extracted from a nearby source-free circular region with a radius of \(188''\). However, the pile-up effect is found to be non-negligible as indicated by the high count rate of $\sim1.5~$counts~s$^{-1}$. To mitigate the pile-up effects, photons within an inner radius of \(8''\) were excluded. For the extracted X-ray spectra, we employed \texttt{xrtmkarf} to build the ancillary response files (ARF). The response matrix functions (RMFs) were then determined using the \texttt{quzcif} tool.

The first {\it Swift}/XRT observation provides a source position, which is enhanced by applying correction of the {\it Swift}/UVOT photometry \citep{Evans2009}: R.A. (J2000.0) = $18^{\rm h}27^{\rm m}30.1^{\rm s}$ and Decl. (J2000.0)= $-9^{\circ}56'41.4''$ (uncertainty of $2.1''$ at the 90\% confidence level). The XRT position is consistent with the FXT position with a separation of $8''$ (note that the uncertainty of FXT position is $10''$). We adopt the XRT coordinates as the X-ray position of EP J182730.0-095633.

\subsubsection{NuSTAR}
\label{sec:nustar}
\textit{NuSTAR} carried out an observation on September 11th, 2024, with a total exposure time of approximately 20.7 ks (PI: Alessio Marino). The data were processed using the standard tools in the \texttt{NuSTARDAS} package. The cleaned level-2 event products were extracted using the \texttt{nupipeline} routine. The source photons were extracted using a circular region with a radius of \(60''\) centered on the source, while the background events were extracted from an annulus region with an inner radius of \(90''\) and an outer radius of \(120''\). Finally, we used \texttt{nuproducts} to extract spectra and light curves for further analysis, and corrected the photon arrival times to the Solar System barycenter using the \texttt{barycorr} tool.

\subsection{Searches for periodic X-ray signals}
\label{sec:xray_periodic_search}
We searched for significant peaks in the power spectra extracted from the \nicer\ and \nustar\ datasets over the whole energy range using the algorithm described by \cite{Israel1996}. This algorithm is based on a fast Fourier transform and accounts for non-Poissonian noise components in the Leahy-normalized \citep{Leahy1983} power spectra. Our analysis did not reveal any significant peaks in any observations above a 3.5$\sigma$ confidence threshold, calculated by considering the number of frequency trials examined. Focusing on the \nicer\ datasets, the most stringent 3-$\sigma$ upper limit on the flux pulsed fraction (which we define as the semi-amplitude of a sinusoidal modulation divided by the mean source count rate) was obtained from the second observation, the one with the longest exposure. This upper limit ranges between 6--10\% for frequencies spanning 30--1000\,Hz. For the \nustar\ data, after combining datasets from both FPMs, the upper limit is between 8--13\% within the same frequency range.

We also searched for periodic signals in the second \nicer\ observation and the \nustar\ observation using Fourier-domain acceleration search techniques. Specifically, we employed the \texttt{accelsearch} pipeline from the \texttt{PRESTO}\footnote{\url{https://github.com/scottransom/presto}} pulsar timing software package \citep{Ransom_2002} to search for signals within the frequency range of 1--1000\,Hz, summing up to eight harmonics. To account for potential power drifts in the Fourier domain, we allowed the powers of signals to drift by up to 200 frequency bins. Moreover, we conducted a ``jerk'' search, allowing the powers of the signals to drift by up to 600 frequency derivative bins, as described by \cite{Andersen2018}. No promising signals were detected.

For \textit{EP}/FXT observations, we also searched the pulsed signal using {\sl powspec} and further checked the signal around the QPO frequency with the blind search technique developed by the Fermi-LAT collaboration \citep{2006ApJ...652L..49A}. No confirmed signals were detected, consistent with the result obtained from {\sl NICER} and {\sl NuSTAR} data. In the end, the upper limit for the FXT timing mode observation is between 6--13\% for frequencies spanning 30--1000\,Hz using the algorithm described by \cite{Israel1996}.

\subsection{Optical and infrared}
\label{sec:optical_datareduction}

\subsubsection{GROND}
\label{sec:grond_datareduction}
EP J182730.0-095633 was observed with the Gamma-ray Burst Optical Near-infrared Detector \citep[GROND,][]{GROND2008};
 mounted at the MPG 2.2\ m telescope at ESO’s La Silla observatory for six nights from September 7th to October 1st 2024 (see Table~\ref{tab:EP J182730.0-095633_opt_infrared_followups}. Observations were performed simultaneously in the $g^{\prime}$, $r^{\prime}$, $i^{\prime}$, $J$, $H$, and $K_{s}$ bands with typical exposures of 30\,min per band and observation. The data were reduced using the standard IRAF-based
GROND pipeline \citep{kruehler2008}. The photometry was calibrated against PanSTARSS DR1 \citep{Chambers2016} ($g^{\prime}$, $r^{\prime}$, $i^{\prime}$) and the Two Micron All Sky
Survey catalogue \citep[2MASS;][]{2MASS2006} ($J$, $H$, and $K_{s}$) and converted into the AB system.

\subsubsection{\textit{SVOM/VT}}
\label{sec:svomvt_datareduction}
The VT (Visible Telescope) is an optical telescope onboard the \textit{Space Variable Objects Monitor mission} \citep[\textit{SVOM,}][]{2016SVOM}. The effective aperture is 43 cm. The field of view is \(26'\)$\times$\(26'\), giving the pixel scale of \(0.76''\). It conducts the observation with two broad-band channels $VT_{\rm B}$ and $VT_{\rm R}$ simultaneously, covering 400--650 nm and 650--1000 nm, respectively. Detailed information on VT can be referred to Qiu et al. (2025, in preparation). EP J182730.0-095633 was observed twice by VT during its commissioning phase via Target Of Oppotunity (ToO) mode, one on 13:23:58 UT, 2024-09-06 and the other on 21:45:55 UT, 2024-09-28. The exposure time was set to be 20\,s for single frame. All the data were processed in a standard manner, including zero correction, dark correction, and flat-field correction. After pre-processing, the images for each band obtained during each observation were stacked to increase the signal-to-noise ratio.

\subsubsection{NOT}
\label{sec:not_datareduction}
The optical photometry follow-up observation was carried out using the 2.56 m Nordic Optical Telescope (NOT; Roque de los Muchachos observatory, La Palma, Spain) equipped with Alhambra Faint Object Spectrograph and Camera (ALFOSC) at 20:42:54 UT on 2024-09-07, i.e. 3.4 days after the trigger, and 4$\times$180\,s frames were obtained in the SDSS-$r$ band. The NOT data are reduced by standard procedures with the \textsc{Image Reduction and Analysis Facility} (\textsc{IRAF}) v2.16 \citep{1986IRAF}, including bias subtraction, flat-field correction and images combination. 

\subsubsection{TRT}
\label{sec:trt_datareduction}
We began to observe EP J182730.0-095633 using the 0.7-m telescope of the Thai Robotic Telescope (TRT) network, located at Sierra Remote Observatories, USA. The observation started at 02:34:30 UT on 2024-09-08, i.e. 3.7 days after the trigger and several frames were obtained in the Johnson $B$ band. Raw data were processed automatically using the TRT's standard data reduction pipeline. We stack clean science frames using \textsc{IRAF} to improve the signal-to-noise ratio. 

\subsubsection{ALT100C}
\label{sec:alt100c_datareduction}
ALT100C is a 1-meter telescope located at Altay observatory, Xinjiang, China. The observation of EP J182730.0-095633 was conducted at 14:49:20 UT on 2024-09-08, i.e. 4.2 days after the trigger in the $g$, $r$ and $i$ bands. The data are also reduced by \textsc{IRAF} and calibrated with nearby PanSTARRS stars. 

\subsubsection{TNT}
\label{sec:tnt_telescope}
The Tsinghua-NAOC Telescope \citep[TNT,][]{Huang2012TNT}, an 80-cm Cassegrain reflecting telescope located at Xinglong Observatory of the National Astronomical Observatories of China (NAOC), was used to observe the field of the X-ray transient EP J182730.0-095633. Observations were conducted in the $r$ and $i$ bands, obtaining $6\times600$\,s exposures per band, with a median observation date of 2024-10-11T11:33:50. After the bias and flat-field correction, the images were stacked by the imcombine program in IRAF. The photometric measurements were carried out using the \textsc{Source Extractor} \citep[\textsc{SExtractor, }][]{1996SourceExtractor}. The automatic aperture photometry was derived from Kron's ``first moment" algorithm \citep{Kron1980}. The instrumental magnitude was then calibrated to Pan-STARR DR1. 

\subsubsection{Mephisto}
\label{sec:mephisto_datareduction}

The Multi-channel Photometric Survey Telescope (Mephisto\footnote{\url{http://www.mephisto.ynu.edu.cn/about/Mephisto}}) is a wide-field ground based telescope with a 1.6-meter primary mirror featuring a field of view of 2 square degrees \citep{2020Mephisto}. 
Observations were conducted at 16:11:28 UT on 2024-09-07, i.e. 3.2 days after the trigger, in the $v$, $r$ and $z$ bands. For data reduction, the raw frames were processed through a dedicated pre-processing pipeline developed for Mephisto. This pipeline includes bias subtraction, dark subtraction, flat fielding, and cosmic ray removal. Photometric calibration was performed using the Gaia BP/RP low-resolution spectra (XP spectra) of non-variable stars. Since there were no detections in either the single frames or the stacked frames for each band, we calculated the $5\sigma$ limiting magnitudes for each frame based on the image's Full Width at Half Maximum (FWHM) and the background fluctuations at the target's position.

\subsection{Radio}
\label{sec:radio_followup}

\subsubsection{ATCA}
Two radio observations were carried out using the Australia Telescope Compact Array (\textit{ATCA}) at 4\,cm, one on September 12th, 2024 from 09:00 to 14:00 UT \citep{2024ATCA_ATEL}, and the other on September 21st, 2024 from 07:30 to 12:30 UT. 
The \textit{ATCA} observations were conducted as part of the project C3615 (PI: Yanan Wang). The central frequencies were set at 5.5\,GHz and 9\,GHz, each with a bandwidth of 2\,GHz. These observations were coordinated with \textit{EP}/FXT observations to investigate any correlations between the radio and X-ray emissions. Data were processed using the \texttt{Common Astronomy Software Applications} \citep[\texttt{CASA};][]{2007ASPC..376..127M, 2022CASA} with standard interferometric imaging methods. Flux density and gain calibrators 1934-638 and 1829-106 were used for the calibration.

\subsubsection{MeerKAT}
We observed the position of EP J182730.0-095633 with the MeerKAT radio telescope \citep{Camilo2018, Jonas2018}, as part of program SCI-20230907-JB-01 (PIs Bright \& Carotenuto). We conducted three observations, each with the same total on-source time of 42 minutes. We observed quasi-simultaneously at L- and S-band, the first at a central frequency of 1.28\, GHz and a total bandwidth of 856 MHz, and the second at 3.06\,GHz (S-band, S4), with a total bandwidth of 875\,MHz. The first observation started on September 21st 2024 at 15:16 UTC (L-band) and 16:51 UTC (S-band). The second observation was performed on October 21 th 2024 at 13:45 UTC (S-band) and on October 22nd 2024 at 19:09 UTC (L-band), while the third observation was conducted on October 28th 2025 at 16:44 UTC (S-band) and 18:14 UTC (L-band). PKS~J1939--6342 and PKS 1830-211 were used as flux and complex gain calibrators, respectively. The data were reduced with the \texttt{OxKAT} pipeline \citep{oxkat}, which performs standard flagging, calibration and imaging using \texttt{tricolour} \citep{Hugo_2022}, \texttt{CASA} and \texttt{WSCLEAN} \citep{Offringa_wsclean}, respectively. In the imaging step, we adopted a Briggs weighting scheme with a $-0.3$ robust parameter. The typical values for the image rms-noise are $\sim$$10\, \mu \rm{Jy}\,\rm{beam}^{-1}$ at S-band and $20 \sim 30 \ \mu \rm{Jy}\,\rm{beam}^{-1}$ at L-band, due to the presence of diffuse emission in the field, stronger at lower frequencies. 

\counterwithin{table}{section}
\counterwithin{figure}{section}
\section{Supplementary material}

In this section, we present the detailed results of the X-ray spectral fitting, optical and infrared follow-up observations and PSDs of \textit{EP}/FXT and \textit{NICER} during the X-ray outburst.

\begin{deluxetable}{lcccccccc}
\tablecaption{
\label{tab:xray_spectralfit}
X-ray spectral fitting results of EP J182730.0-095633.}
\centering
\tablehead{
\colhead{Obs date} 
& \colhead{$T-T_0$}  
& \colhead{Instrument}
& \colhead{OBSID}
& \colhead{$N_{\rm H}$}
& \colhead{$\Gamma$}
& \colhead{$F_{\rm unabsorbed, 0.5-10}$}
& \colhead{${\chi^2}/{\rm d.o.f.}$}
& \colhead{Exposure}\\
& \colhead{(day)} 
& \colhead{}  
& \colhead{}
& \colhead{$(10^{22}~{\rm cm^{-2}})$}
& \colhead{}
& \colhead{(${\times 10^{-11}~{\rm erg~s^{-1}~cm^{-2}}}$)}
& \colhead{}
& \colhead{(seconds)}}
\startdata
2024-09-01/02/03  & $-1.8$ & EP/WXT & 06800000061/63/68 & $3.1*$ & 2.0* & $<4.3$ & \nodata & $27230$ \\
2024-09-04  & $0$  & EP/WXT & 06800000067 & $3.1*$ & $2.0_{-1.2}^{+1.3}$ & $14_{-5}^{+10}$ & $58.3/68$ & $6075$ \\
          &   $0.3$ &  EP/WXT      & 08500000148 & $3.1*$ & $1.2_{-0.7}^{+0.7}$ & $37_{-9}^{+16}$ & $99.8/104$ & $9246$ \\
2024-09-05  & $1.0$  & EP/WXT & 06800000071 & $3.1*$ & $2.6_{-0.7}^{+0.7}$ & $52_{-15}^{+29}$ & $50.1/47$ & $2987$ \\
          &   $1.3$ &  EP/WXT      & 06800000069 & $3.1*$ & $1.6_{-0.4}^{+0.4}$ & $40_{-6}^{+7}$ & $130.5/146$ & $9155$ \\
2024-09-06  & $2.1$  & EP/FXT & 06800000074 & $3.1_{-0.1}^{+0.1}$ & $2.00_{-0.08}^{+0.08}$ & $35_{-2}^{+2}$ & $307.8/297$ & $5966$ \\
          &   $2.4$ & EP/WXT & 06800000070 & $3.1*$ & $1.9_{-0.6}^{+0.7}$ & $30_{-7}^{+8}$ & $129.5/94$ & $6070$ \\
2024-09-07  & $3.3$  & EP/FXT & 06800000081 & $3.3_{-0.2}^{+0.2}$ & $2.03_{-0.10}^{+0.10}$ & $23_{-1}^{+2}$ & $195.6/189$ & $2983$ \\
          &  $3.3$  & NICER/XTI  & 7204880101  & $3.2_{-0.2}^{+0.2}$ & $2.1_{-0.17}^{+0.14}$ & $27_{-2}^{+3}$ & $109.39/117$ & $223$ \\
2024-09-08   &  $3.8$  & Swift/XRT  & 00016796001 & $3.4_{-0.5}^{+0.5}$ & $2.02_{-0.21}^{+0.22}$ & $20_{-3}^{+4}$ & $61.5/50$ & $2864$ \\
&  $4.1$  &   NICER/XTI     & 7204880102 & $3.1_{-0.1}^{+0.1}$ & $2.11_{-0.04}^{+0.06}$ & $25.3_{-0.6}^{+0.7}$ & $607.69/603$ & $3921$ \\
             & $4.1$  & EP/FXT & 06800000079 & $3.3_{-0.1}^{+0.1}$ & $2.01_{-0.07}^{+0.07}$ & $21.0_{-0.9}^{+1.0}$ & $303.7/276$ & $5941$ \\
            &  $4.5$  &  EP/FXT  & 06800000080 & $3.3_{-0.1}^{+0.2}$ & $2.04_{-0.09}^{+0.09}$ & $20_{-1}^{+1}$ & $221.8/209$ & $3935$ \\
2024-09-09  & $5.2$  & EP/FXT & 06800000082 & $3.2_{-0.2}^{+0.2}$ & $2.06_{-0.11}^{+0.11}$ & $18_{-1}^{+1}$ & $145.3/151$ & $2984$ \\
          &  $5.5$  &  EP/FXT  & 06800000083 & $3.4_{-0.2}^{+0.2}$ & $2.10_{-0.13}^{+0.13}$ & $17_{-1}^{+2}$ & $154.8/128$ & $2716$ \\
          & $4.9$ & NICER/XTI  & 7204880103  & $3.2_{-0.1}^{+0.1}$ & $2.22_{-0.08}^{+0.08}$ & $24_{-1}^{+1}$ & $527.82/515$ & $1543$ \\
2024-09-10 &  $5.9$  & EP/FXT & 06800000087 & $3.1_{-0.2}^{+0.2}$ & $2.04_{-0.09}^{+0.09}$ & $14.9_{-0.8}^{+0.9}$ & $231.7/218$ & $5319$\\
          &  $6.4$  &  EP/FXT    & 06800000088 & $3.3_{-0.3}^{+0.3}$ & $2.15_{-0.19}^{+0.20}$ & $14_{-2}^{+2}$ & $75.5/64$ & $1580$ \\
2024-09-11 & $6.9$  & EP/FXT & 06800000089 & $3.2_{-0.2}^{+0.2}$ & $2.09_{-0.10}^{+0.10}$ & $12.8_{-0.7}^{+0.9}$ & $188.4/185$ & $5361$ \\
          &  $7.2$  & NuSTAR & 91001334002 & $3.7\pm1.3$ & $1.92\pm0.03$ & $8.2\pm0.3$ & $718.7/765$ & $2690$ \\
          &   $7.4$ &   EP/FXT     & 06800000090 & $2.9_{-0.3}^{+0.4}$ & $2.00_{-0.20}^{+0.21}$ & $11_{-1}^{+2}$ & $82.5/57$ & $1500$ \\
2024-09-12 & $7.8$  & NICER/XTI & 7204880105 & $3.3_{-0.4}^{+0.6}$ & $2.54_{-0.30}^{+0.47}$ & $15_{-3}^{+6}$ & $168.8/165$ & $703$ \\
    & $7.8$  & EP/FXT & 06800000091 & $3.1_{-0.1}^{+0.2}$ & $2.09_{-0.09}^{+0.09}$ & $10.7_{-0.6}^{+0.7}$ & $240.7/221$ & $7581$ \\
2024-09-13 & $8.9$  & EP/FXT & 06800000095 & $3.0_{-0.2}^{+0.2}$ & $2.18_{-0.12}^{+0.12}$ & $7.6_{-0.6}^{+0.7}$ & $132.5/133$ & $5886$ \\
          &  $9.2$  & EP/FXT  & 06800000096 & $3.0_{-0.2}^{+0.2}$ & $2.07_{-0.12}^{+0.13}$ & $6.8_{-0.5}^{+0.6}$ & $144.3/124$ & $5976$ \\
2024-09-14 & $10.1$ & EP/FXT & 06800000097 & $2.6_{-0.2}^{+0.2}$ & $1.93_{-0.14}^{+0.15}$ & $4.5_{-0.3}^{+0.4}$ & $118.1/97$ & $5978$ \\
2024-09-15 & $11.1$ & EP/FXT & 06800000098 & $2.8_{-0.3}^{+0.3}$ & $2.08_{-0.20}^{+0.22}$ & $3.3_{-0.3}^{+0.5}$  & $39.9/48$ & $4091$ \\
2024-09-17 & $12.9$ & EP/FXT & 06800000110 & $2.4_{-0.6}^{+0.7}$ & $1.93_{-0.42}^{+0.47}$ & $1.5_{-0.2}^{+0.5}$ & $19.4/16$ & $2992$ \\
2024-09-20 & $15.9$ & EP/FXT & 06800000111 & $3.1^{*}$ & $2.07_{-0.25}^{+0.25}$ & $0.49_{-0.05}^{+0.06}$ & $132.7/114$ & $5987$ \\
2024-09-21 & $16.9$ & EP/FXT & 06800000114 & $3.1^{*}$ & $2.30_{-0.31}^{+0.32}$ & $0.24_{-0.03}^{+0.04}$ & $100.1/85$ & $8979$ \\
2024-09-22 & $17.8$ & EP/FXT & 06800000117 &$3.1^{*}$ & $1.26_{-0.81}^{+0.81}$ & $0.059_{-0.002}^{+0.003}$ & $11.4/17$ & $5960$ \\
2024-09-23 & $19.4$ & EP/FXT & 06800000118 & $3.1^{*}$ & $2.0^{*}$ & $<0.068$ & \nodata & $1387$\\
2024-09-24 & $20.0$ & EP/FXT & 06800000120 & $3.1*$ & $1.47_{-1.40}^{+1.40}$ & $0.043_{-0.018}^{+0.045}$& $10.8/14$ & $5842$ \\
          &  $20.6$  &     EP/FXT    & 06800000123 & $3.1*$ & $1.37_{-0.98}^{+0.98}$ & $0.030_{-0.012}^{+0.020}$ & $13.5/12$ & $8095$ \\
2024-11-01 & $57.9$ & Swift/XRT & 00016796002 & $3.1^{*}$ & $2.0^{*}$ & $<0.037$ & ... & $2582$ \\
\enddata
\tablecomments{$T_0$ is set as the starting time of the WXT observation during which the source was firstly detected, at 2024-09-04T10:12:56 (UTC). The parameters marked with an asterisk were held fixed at the quoted values in the fits.}
\end{deluxetable}

\counterwithin{table}{section}
\counterwithin{figure}{section}
\begin{table*}[ht]
\caption{
\label{tab:EP J182730.0-095633_opt_infrared_followups}
Optical and infrared follow-up observations of EP J182730.0-095633.}
\centering
\begin{tabular}{lccccccc}
\hline
Obs date & $T-T_0$ (day) & Instrument & Filter & Detection & Observed magnitude or $5\sigma$ upper limit \\
\hline
2024-09-06 & 2.1 & \textit{SVOM}/VT & ${VT_R}, {VT_B}$ & N &  $>21.4$ / $>22.9$ \\
2024-09-07 & 2.6 & GROND & $J,H,K_{s}$ & Y & 
$17.97\pm0.22^{\rm J}$ / $16.73\pm0.13^{\rm H}$ /  
$16.32\pm0.13^{\rm K}$  \\
& & & $g^{\prime}$, $r^{\prime}$, $i^{\prime}$ & N & $\textgreater23^{\rm g}$ / $\textgreater22.6^{\rm r}$ / $\textgreater21.5^{\rm i}$  \\
   & 3.2 & Mephisto & $v,r,z$ & N & $\textgreater20.7^{\rm v}$ / $\textgreater20.5^{\rm r}$ / $\textgreater19.36^{\rm z}$\\
   & 3.4 & NOT & $r$ & Y & $23.87\pm0.35^{\rm r}$\\   
2024-09-08 & 3.6 & GROND & $J,H,K_{s}$ & Y & $18.09\pm0.18^{\rm J}$ / $16.81\pm0.18^{\rm H}$ /  
$16.43\pm0.16^{\rm K}$  \\
& & & $g^{\prime}$, $r^{\prime}$, $i^{\prime}$ & Y & $\textgreater23.9^{\rm g}$ / $23.86\pm0.30^{\rm r}$ / $21.77\pm0.10^{\rm i}$  \\
      & 3.7 & TRT & $B$ & N &  $\textgreater21.4^{\rm B}$ \\
   & 4.2 & ALT100C & $r, g, i$ & N & $\textgreater22.3^{\rm r}$ / $\textgreater21.7^{\rm g}$ /  $\textgreater20.9^{\rm i}$ \\
2024-09-09 & 4.6 & GROND & $J,H,K_{s}$ & Y & $18.01\pm0.18^{\rm J}$ / $16.93\pm0.16^{\rm H}$ /  $16.50\pm0.15^{\rm K}$  \\
& & & $g^{\prime}$, $r^{\prime}$, $i^{\prime}$ & Y & $\textgreater23.6^{\rm g}$ / $\textgreater23.2^{\rm r}$ / $21.95\pm0.15^{\rm i}$  \\
2024-09-11 & 6.7 & GROND & $J,H,K_{s}$ & N & - / - /  
-  \\
& & & $g^{\prime}$, $r^{\prime}$, $i^{\prime}$ & N & $\textgreater21.2^{\rm g}$ / $\textgreater21.5^{\rm r}$ / $\textgreater20.7^{\rm i}$  \\
2024-09-18 & 13.7 & GROND & $J,H,K_{s}$ & N & $\textgreater19.2^{\rm g}$ / $\textgreater17.7^{\rm r}$ / $\textgreater17.4^{\rm i}$  \\
& & & $g^{\prime}$, $r^{\prime}$, $i^{\prime}$ & N & $\textgreater21.7^{\rm g}$ / $\textgreater22.3^{\rm r}$ / $\textgreater21.6^{\rm i}$  \\
2024-09-28 & 24.5 & \textit{SVOM}/VT & ${VT_R}$, ${VT_B}$ & N & $>21.6$ / $>22.3$ \\
2024-10-01 & 27.6 & GROND & $J,H,K_{s}$ & N & - / - /  
-  \\
& & & $g^{\prime}$, $r^{\prime}$, $i^{\prime}$ & N & $\textgreater21.0^{\rm g}$ / $\textgreater21.2^{\rm r}$ / $\textgreater21.0^{\rm i}$  \\
2024-10-11 & 37.0 & TNT & $r, i$ & N & $\textgreater18.24^{\rm r}$ / $\textgreater17.54^{\rm i}$ \\
\hline
\end{tabular}
\tablecomments{$T_0$ is set as the starting time of the WXT observation during which the source was firstly detected, at 2024-09-04T10:12:56 (UTC). With poor seeing conditions during the GROND observations in the $J$, $H$, and $K$ bands on September 11th and October 1st, it was not feasible to accurately constrain the flux of the source due to its proximity to other brighter sources. As a result, the magnitudes in these bands are not provided.}
\end{table*}

\counterwithin{table}{section}
\counterwithin{figure}{section}
\begin{table*}[ht]
\caption{
\label{tab:EP J182730.0-095633_radio_followups}
Radio follow-up observations of EP J182730.0-095633.}
\centering
\begin{tabular}{lccccccc}
\hline
Obs date & $T-T_0$ (day) & Instrument & Filter & Detection & Peak flux density / $3\sigma$ upper limit \\
& & & & & $\mu {\rm Jy/beam}$ \\
\hline
2024-09-12 & $7.9$ & \textit{ATCA} & 5.5 GHz & Y & $101.2\pm5.4$ \\
 & $7.9$ & \textit{ATCA} & 9 GHz & Y & $126.3\pm5.2$ \\
2024-09-21 & $16.9$ & \textit{ATCA} & 5.5 GHz & N & $<57$ \\
 & $16.9$ & \textit{ATCA} & 9 GHz & N & $<60$ \\
 & $17.2$ & \textit{MeerKAT} & 1.28 GHz (L-band) & N & $<69$  \\
 & $17.3$ & \textit{MeerKAT} & 3.06 GHz (S-band) & Y & $33\pm4$\\
2024-10-21 & $47.1$ & \textit{MeerKAT} & 3.06 GHz (S-band) & N & $<36$  \\ 
2024-10-21 & $48.4$& \textit{MeerKAT} & 1.28 GHz (L-band) & Y & $58\pm12$ \\ 
2024-10-22 & $54.3$ & \textit{MeerKAT} & 3.06 GHz (S-band) & N & $<36$   \\
2024-10-22 & $54.3$ & \textit{MeerKAT} & 1.28 GHz (L-band) & N & $<105$ \\

\hline
\end{tabular}
\tablecomments{$T_0$ is set as the starting time of the WXT observation during which the source was firstly detected, at 2024-09-04T10:12:56 (UTC).}
\end{table*}

\begin{figure*}[ht]
    \centering
    \includegraphics[width=1.0\textwidth]{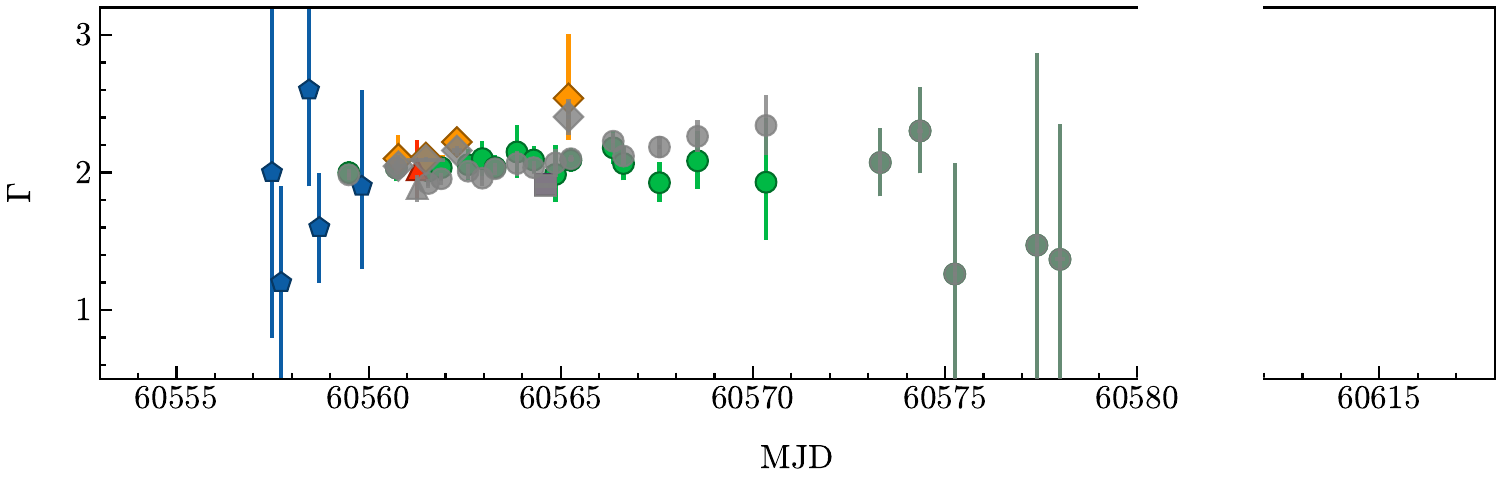}
    \caption{Evolutionary trend of the photon index $\Gamma$ using two different data analysis strategies for \textit{EP}/FXT, \textit{NICER}, \textit{Swift}/XRT and \textit{NuSTAR} data. The result shown as gray symbols is obtained by fixing the column density $N_{\rm H}$ to the typical parameter value of $3.1\times10^{22}~{\rm cm^{-2}}$ during the outburst. The result obtained with variable column density (same as in Figure \ref{fig:xray_evolution}) is also plotted for comparison.}
    \label{fig:gamma_evolution_with_nh_fixed}
\end{figure*}

\begin{figure*}[ht]
    \centering
    \includegraphics[width=1.0\textwidth]{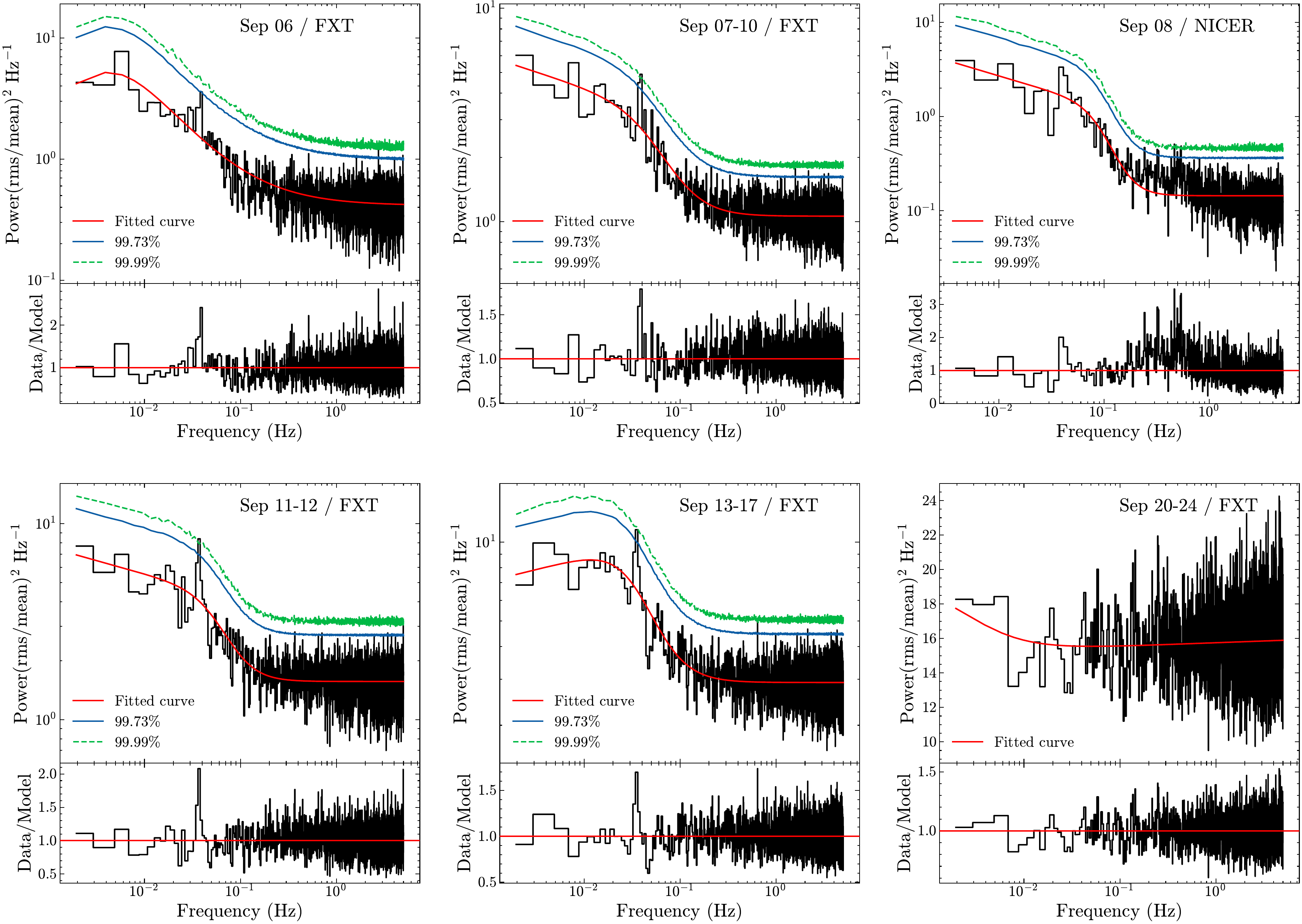}
    \caption{PSDs of \textit{EP}/FXT and \nicer~during the outburst. 
    The centroid frequency of the most prominent QPO holds steady at $\sim$0.04\,Hz. For the \textit{EP}/FXTA (TM mode) data, we combined them to improve the signal-to-noise ratio. The absence of the QPO signal in the final FXT dataset (from September 20th to 24th) is likely attributed to the extremely low count rate of the source during this period.
    }
    \label{fig:FXT_PSD}
\end{figure*}

\begin{figure}[h!]
    \centering
    \includegraphics[width=0.45\linewidth]{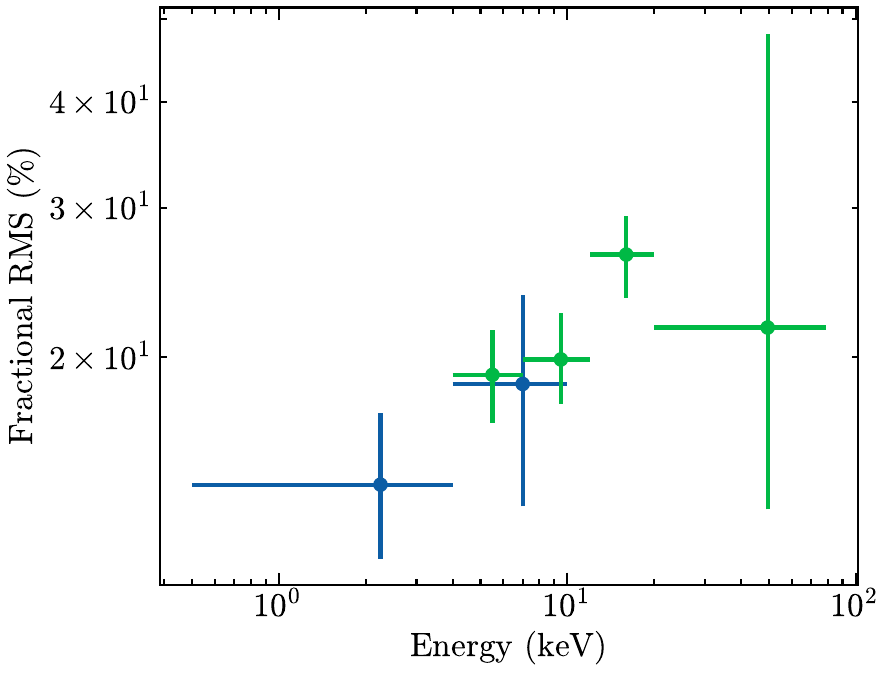}
    \includegraphics[width=0.45\linewidth]{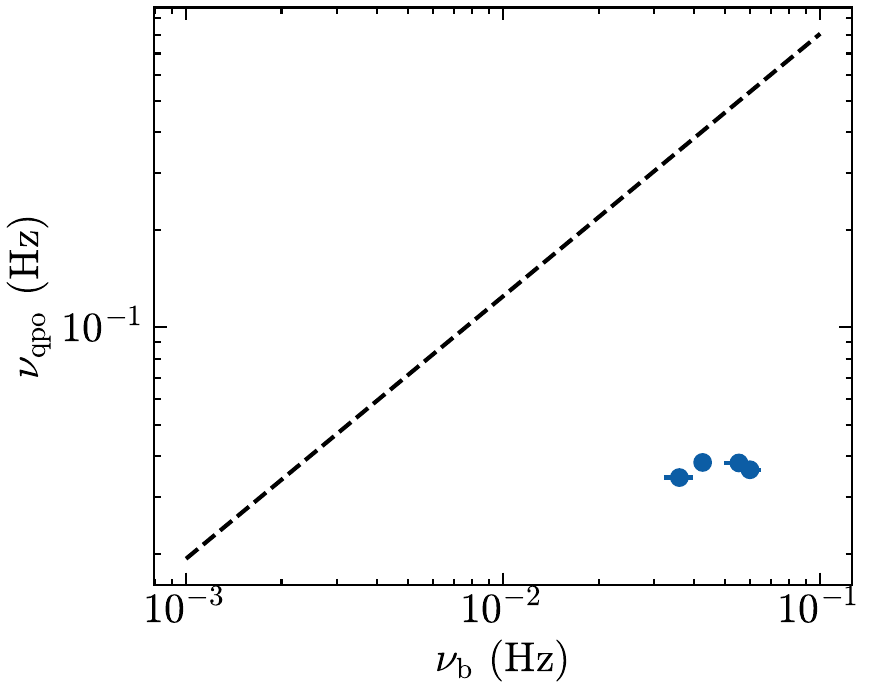}
    \caption{Left panel: the RMS spectrum calculated using the same data set as Figure \ref{fig:spectral_fit} (\textit{EP}/FXT+\textit{NuSTAR} on September 11th, 2024). The blue dots are from \textit{EP-FXT} and green dots from \textit{NuSTAR}.
    Right panel: the relation between the low frequency break ($\nu_{\rm b}$) of the PSD and the centroid frequency of the QPO signal ($\nu_{\rm qpo}$), with the results of EP J182730.0-095633 denoted as blue circles. Here we use the PSDs from Sep 07–10 (FXT), Sep 11 (NuSTAR), Sep 11–12 (FXT), and Sep 13–17 (FXT), where the QPO significance exceeds $3\sigma$. The black dashed line is the empirical W-K relation between the two variables for type-C LFQPOs. \(\nu_{\rm qpo}\) are \(38.2 \pm 0.2\), \(38.3 \pm 0.3\), \(36.4 \pm 0.4\), and \(34.4 \pm 0.4\)~mHz for the four PSDs, respectively, while \(\nu_{\rm b}\) are \(55 \pm 6\), \(43 \pm 2\), \(60 \pm 5\), and \(36 \pm 4\)~mHz.
    }
    \label{fig:qpo_discussion}
\end{figure}

\bibliography{ref}{}

\bibliographystyle{aasjournal}
\end{document}